\documentclass[a4paper,11pt]{article}

\usepackage{iftex}
\usepackage{comment}

\ifluatex
\else
\pdfoutput=1 
\usepackage[utf8]{inputenc}
\usepackage[T1]{fontenc} 
\fi

\usepackage{jheppub} 

\usepackage{physics}
\usepackage{hyperref}

\usepackage{xcolor}
\usepackage{xifthen}
\usepackage{tikz}
\usepackage{standalone}
\input{figures/figure-preamble}
\tikzset{external/mode={list and make}}

\usepackage{subcaption}

\usepackage{amsmath}
\usepackage{amsthm}
\usepackage{amssymb}

\usepackage{bbm}

\usepackage{cleveref}
\usepackage{siunitx}

\newcommand{\ads}{\text{AdS}}
\newcommand{\lgqeham}{H_Q}
\newcommand{\idop}{\mathbbm{1}}

\DeclareMathOperator{\sgn}{sign}
\DeclareMathOperator{\Pf}{Pf}

\title{Rescuing a black hole in the large-$q$ coupled SYK model}
\author[a]{Yuri D. Lensky,}
\author[a]{Xiao-Liang Qi}
\affiliation[a]{\small \em  Stanford Institute for Theoretical Physics, Stanford University, Stanford CA 94305 USA}
\emailAdd{ydl@stanford.edu}
\emailAdd{xlqi@stanford.edu}

\newcommand{\XLQc}[1]{{\color{red} \textbf{[XLQ:} #1\textbf{]}}}
\newcommand{\XLQ}[1]{{\color{purple}#1}}
\providecommand{\autocite}[1]{\cite{#1}}

\abstract{In this paper, we develop a general effective theory for two copies of the Sachdev-Ye-Kitaev (SYK) model with a time-dependent bilinear coupling. For a quantum quench problem with an initial state of the thermofield double state, we show how the evolution of the system is described by a complex reparametrization field with a classical Hamiltonian. We study correlation functions in this system and compare the large-$q$ theory with the bulk low energy effective theory. In particular, we study the special case of a ``rescued black hole'', which describes how a time-evolved thermofield double state can evolve to the ground state of a coupled SYK model by a carefully tuned time-dependent coupling. In the low energy region, there is a holographic dual interpretation, which is a geometry that crosses over from an eternal black hole to a global AdS$_2$ vacuum. This family of geometries allow us to access the bulk region that would be the black hole interior without the rescue process. By comparing the large-$q$ and low energy theory, we find that even in the low energy region the deviation from the low energy theory cannot be neglected if the rescue process starts late. This provides evidence that the low energy effective theory of the bulk fails near the inner horizon of the black hole. We note the possibility of a connection to a two-dimensional analog of the higher-dimensional black hole singularity.}

\begin{document}

\maketitle

\tableofcontents

\section{Introduction}
\label{sec:introduction}

In recent years, progress in holographic duality (also known as the anti-de Sitter/conformal field theory duality (AdS/CFT))\cite{maldacena1999large,witten1998anti,gubser1998gauge} has uncovered many deep connections between quantum gravity and quantum many-body physics. For example, the ground state of a CFT is dual to the AdS vacuum, while a thermal state is dual to a black hole. Despite a more and more complete understanding of the duality between bulk and boundary physics, there are still many open questions about the black hole interior. In principle, operators in the interior can be reconstructed from the boundary, but it is unclear how the geometry and the dynamics of quantum fields in the interior region are determined by the boundary physics. In particular, what is the boundary interpretation of the black hole singularity? (For a recent work related to this question see \cite{grinberg2020proper}.) 

\begin{figure}[htb!]
  \centering
  \begin{subfigure}{0.32\linewidth}
    \centering
    \begin{tikzpicture}
  \def\boundrho{2.4}
  \def\toprindlerextent{0.001}
  \def\bottomrindlerextent{0.001}
  \pgfmathsetmacro\coshboundrho{0.5*(e^\boundrho + e^(-\boundrho))}
  \pgfmathsetmacro\fermiontG{pi/2-0.4}
  \pgfmathsetmacro\fermionsigma{-acos(cos(\fermiontG r) / \coshboundrho) * pi / 180}
  \draw[conformal boundary]
  (-pi/2, -pi/2) -- (-pi/2, 3*pi/2);
  \draw[conformal boundary]
  (pi/2, -pi/2) -- (pi/2, 3*pi/2);
  \draw[syk boundary] plot[smooth, domain=(-pi/2+\bottomrindlerextent):(pi/2-\toprindlerextent)]
  ({acos(cos(\x r) / \coshboundrho) * pi / 180}, \x);
  \draw[syk boundary] plot[smooth, domain=(-pi/2+\bottomrindlerextent):(pi/2-\toprindlerextent)]
  ({-acos(cos(\x r) / \coshboundrho) * pi / 180}, \x);

  \path[pattern=crosshatch, pattern color=gray!30] (0, 0) -- (pi/2, pi/2) -- (0, pi) -- (-pi/2, pi/2) -- cycle;

  \draw[light ray marker] (-pi/2, -pi/2) -- (pi/2,pi/2);
  \draw[light ray marker] (pi/2, -pi/2) -- (-pi/2,pi/2);
  \draw[light ray marker] (-pi/2,pi/2) -- ++(pi,pi);
  \draw[light ray marker] (pi/2,pi/2) -- ++(-pi,pi);

  \begin{scope}
    \clip (-pi/2, -pi/2) rectangle (pi/2, 3*pi/2);
    \draw[fermion propagator] (\fermionsigma, \fermiontG) -- ++(pi, pi);
    \draw[boundary insertion point, fill=black] (\fermionsigma, \fermiontG) circle;
  \end{scope}
\end{tikzpicture}
    \caption{``Thermal'' or ``decoupled''.}
    \label{fig:rindler-sol-ads-embedding}
  \end{subfigure}
  \begin{subfigure}{0.32\linewidth}
    \centering
    \begin{tikzpicture}
  \def\globalsigmad{0.07}
  \pgfmathsetmacro\fermiontG{pi/2-0.4}

  \draw[conformal boundary]
  (-pi/2, -pi/2) -- (-pi/2, 3*pi/2);
  \draw[conformal boundary]
  (pi/2, -pi/2) -- (pi/2, 3*pi/2);

  \path[pattern=crosshatch, pattern color=gray!30] (0, 0) -- (pi/2, pi/2) -- (0, pi) -- (-pi/2, pi/2) -- cycle;

  \draw[light ray marker] (-pi/2, -pi/2) -- (pi/2,pi/2);
  \draw[light ray marker] (pi/2, -pi/2) -- (-pi/2,pi/2);
  \draw[light ray marker] (-pi/2,pi/2) -- ++(pi,pi);
  \draw[light ray marker] (pi/2,pi/2) -- ++(-pi,pi);

  \draw[syk boundary] (-pi/2 + \globalsigmad, -pi/2) -- ++(0, 2*pi);
  \draw[syk boundary, name path=right boundary] (pi/2 - \globalsigmad, -pi/2) -- ++(0, 2*pi);

  \begin{scope}[overlay]
    \path[name path=bg prop]
    (-pi/2 + \globalsigmad, \fermiontG) coordinate (prop start)
    -- ++(pi, pi);
    \draw[fermion propagator,
    name intersections={of=bg prop and right boundary}]
    (prop start) -- (intersection-1) coordinate (prop end);
    \draw[boundary insertion point, fill=black] (prop start) circle;
    \draw[boundary insertion point, fill=black] (prop end) circle;
  \end{scope}
\end{tikzpicture}
    \caption{``Global'' or ``fixed point''.}
    \label{fig:global-sol-ads-embedding}
  \end{subfigure}
  \begin{subfigure}{0.32\linewidth}
    \centering
    \begin{tikzpicture}[scale=1]
  \def\boundrho{2.4}
  \def\toprindlerextent{0.1}
  \def\bottomrindlerextent{0.001}
  \def\uzinsertiongap{0.3}
  \def\u1insertiongap{0.5}
  \pgfmathsetmacro\coshboundrho{0.5*(e^\boundrho + e^(-\boundrho))}
  \tikzset{declare function={
    rightrindlertG(\tG) = acos(cos(\tG r) / \coshboundrho) * pi / 180;
    leftrindlertG(\tG) = -rightrindlertG(\tG);
  }}
  \draw[conformal boundary]
  (-pi/2, -pi/2) -- (-pi/2, 3*pi/2);
  \draw[conformal boundary]
  (pi/2, -pi/2) -- (pi/2, 3*pi/2);
  \path[pattern=crosshatch, pattern color=gray!30] (0, 0) -- (pi/2, pi/2) -- (0, pi) -- (-pi/2, pi/2) -- cycle;

  \draw[light ray marker] (-pi/2, -pi/2) -- (pi/2,pi/2);
  \draw[light ray marker] (pi/2, -pi/2) -- (-pi/2,pi/2);
  \draw[light ray marker] (-pi/2,pi/2) -- ++(pi,pi);
  \draw[light ray marker] (pi/2,pi/2) -- ++(-pi,pi);

  \draw[syk boundary, name path=right rindler boundary] plot[smooth, domain=(-pi/2+\bottomrindlerextent):(pi/2-\toprindlerextent)]
  ({rightrindlertG(\x)}, \x) coordinate (right rindler end);
  \draw[syk boundary, name path=left rindler boundary] plot[smooth, domain=(-pi/2+\bottomrindlerextent):(pi/2-\toprindlerextent)]
  ({leftrindlertG(\x)}, \x) coordinate (left rindler end);

  \foreach \rindend in {left rindler end, right rindler end}
  {
    \path let \p1 = (\rindend) in [name path=right global boundary] (\p1) -- (\x1, 3*pi/2);
  }

  \begin{scope}[overlay]
    \coordinate (left u1 insertion) at ({leftrindlertG(pi/2-\u1insertiongap)}, pi/2-\u1insertiongap);
    \path[name path=bg u1 prop] (left u1 insertion) -- ++(pi, pi);
    \path[name intersections={of=bg u1 prop and right global boundary}]
    (intersection-1) coordinate (right u2 insertion);
  \end{scope}

  \draw[fermion propagator] (left u1 insertion) -- (right u2 insertion);

  \foreach \rindend in {left rindler end, right rindler end}
  {
    \path let \p1 = (\rindend) in [draw, syk boundary, color=red] (\p1) -- (\x1, 3*pi/2);
  }

  \draw[boundary insertion point, fill=black] (left u1 insertion) circle;
  \draw[boundary insertion point, fill=black] (right u2 insertion) circle;
\end{tikzpicture}
    \caption{``Rescued''.}
    \label{fig:rescued-sol-ads-embedding}
  \end{subfigure}
  \caption{Embeddings in $\ads_2$ of various solutions of the coupled JT gravity theory. These can also be thought of as dual geometries for the coupled SYK dots considered in this paper. The dotted lines are the conformal boundary (spatial infinity in $\ads_2$) and the solid lines are the physical boundary at a finite location. (a) is an eternal black hole solution corresponding to the thermofield double state of the SYK model with decoupled time-evolution. (b) is the geometry with translation symmetry in global time, corresponding to the ground state of a coupled SYK model. (c) is a geometry interpolating between the two previous ones, obtained by a fine-tuned time-dependent coupling.}
  \label{fig:basic-ads2-embedded-geometries}
\end{figure}
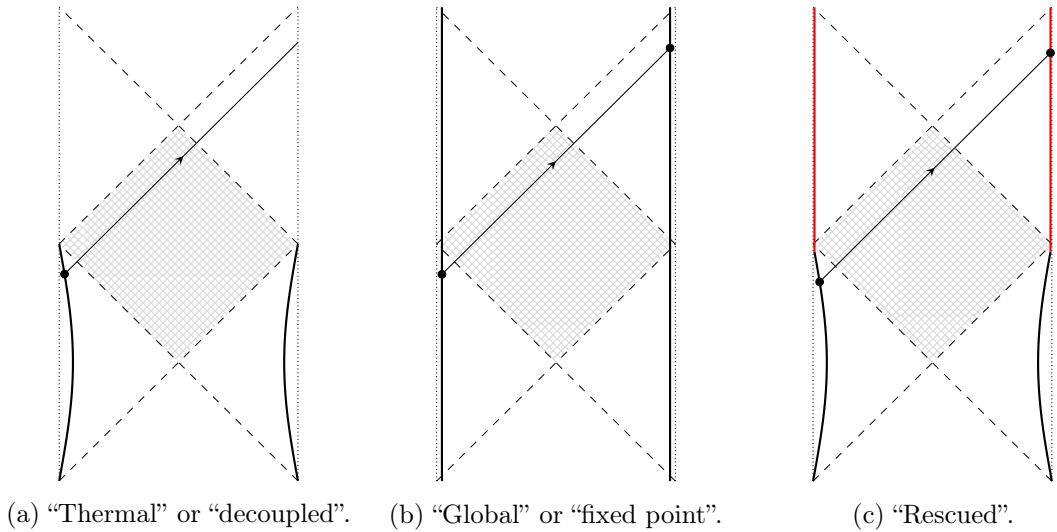

Motivated by this question, in this paper we make an attempt to better understand the interior of a two-dimensional black hole, by studying the dual theory of the Sachdev-Ye-Kitaev (SYK) model\cite{sachdev1993gapless,kitaev2014hidden,kitaev2015simple,maldacena2016conformal,maldacena2016remarks}. The SYK model describes $N$ coupled Majorana fermions in 1d, which in the large-$N$, low energy limit is proposed to be dual to 2d Jackiw-Teitelboim gravity\cite{jackiw1985lower,teitelboim1983gravitation} coupled with matter. Different states in the SYK model are dual to different boundary configurations of the JT gravity, which are the only gravitational degrees of freedom in this system. The thermofield double state is dual to an eternal black hole geometry, as is illustrated in Figure~\ref{fig:rindler-sol-ads-embedding}. Although there is no curvature singularity in 2d, there is an analog which is the contour of zero dilaton value (which is the location of singularity if we obtain JT gravity from dimensional reduction of a higher dimensional black hole). From the SYK model point of view it is unclear what this means, but the natural question is whether the ``semi-classical gravity + matter'' description breaks down somewhere in the interior (hatched region in Fig. \ref{fig:rindler-sol-ads-embedding}). The difficulty with studying the interior region is that, by definition, signal from the interior cannot be sent to the boundary. For example, a scattering event may occur in the interior, but it is unclear how the boundary learns about it. A useful tool for exploring the interior region is the Gao-Jafferis-Wall proposal\cite{gao2017traversable} (see also \cite{maldacena2017diving}) of creating a traversable wormhole by turning on relevant couplings between the two boundaries. In the traversable wormhole geometry, the physics in the used-to-be-interior region now becomes accessible from the boundary. For the SYK model, Ref. \cite{maldacena2018eternal} showed that the wormhole can be made traversable for arbitrarily long time by turning on a particular fermion bilinear coupling between two SYK islands. In particular, the ground state of such a coupled SYK model is dual to portion of the global AdS$_2$ geometry with global time translation invariant boundaries (shown in Figure~\ref{fig:global-sol-ads-embedding}). This is also called an eternal traversable wormhole geometry. In this case the two boundaries are in causal contact, and there are no inaccessible regions in the bulk. 

In this paper, we would like to make use of the eternal traversable wormhole geometry as a tool to probe the black hole interior. We consider the ``rescued black hole geometry'' shown in Figure~\ref{fig:rescued-sol-ads-embedding}, which interpolates between a two-sided black hole geometry (corresponding to two decoupled SYK models in an entangled state) and an eternal wormhole geometry. The interpolation starts from a finite rescue time $u_{r-}$. As $u_{r-}$ gets later and later, the geometry up to the inner horizon (upper edge of the shaded region in the figure) is closer and closer to a black hole geometry. On the other hand, the entire region remains accessible from the boundary. Although there is no singularity in the rescued black hole geometry, we are interested in seeing residual effects of the ``would-be-formed'' singularity if the black hole were not rescued. More precisely, this means looking for signatures that the low energy effective description of semiclassical gravity plus matter starts to fail in some part of the interior region.

For this purpose, we generalize the results of Ref. \cite{maldacena2018eternal} and develop a general theory of the finite energy coupled SYK model in the large-$q$ limit. The large-$q$ limit allows us to go beyond the low energy effective theory description and directly carry UV complete calculations, so that nontrivial comparisons with the low energy JT gravity results can be made. We show that the large-$N$ large-$q$ dynamics of the SYK model with a generic time-dependent coupling and a thermofield double initial state can be described by a complex reparametrization field, which is a generalization of the boundary location in AdS$_2$. The dynamics of the complex reparametrization field is described by the canonical equations of motion of a Hamiltonian, which reduces to the Schwarzian dynamics in the low energy limit. This new Hamiltonian description allows us to study two-point functions and certain four-point functions at finite energy, and compare with the low energy Schwarzian theory. We find interesting finite energy effects, such as a shift of the gravitational excitation frequency relative to the conformal matter excitation frequency. We study the rescued black hole geometry, and show that even if the initial state is a low temperature thermofield double state, and the couplings are kept small, there are certain four-point functions for which the low energy description in terms of free matter propagating on $\ads_2$ is wrong by an order-$1$ fraction if the rescue process is turned on at a late time $u_{r-}$. This suggests that the bulk physics near the inner horizon region is indeed distinct from the predictions of the low energy theory, {\it i.e.} JT gravity coupled with matter.

The remainder of this work is organized as follows. In Section~\ref{sec:coupled-syks-at}, we describe the mapping of the large-$q$ equations to a problem of determining a single complex reparametrization of time governed by a simple Hamiltonian, and give an overview of the solutions. In Section~\ref{sec:bulk-interpretations}, we show how the large-$q$ theory reduces to the coupled Schwarzian effective theory in the low energy limit, and propose a mapping of solutions of the large-\(q\) theory to a pair of AdS boundaries. 
As a first application, in Section~\ref{sec:four-point-functions} we study correlation functions and highlight some simple finite coupling corrections to known results. In Section~\ref{sec:rescue}, we study the ``rescued'' geometry shown in Figure~\ref{fig:rescued-sol-ads-embedding} in detail. We explain how to reach this geometry, and discuss some four-point functions in this setting. We highlight the indications that corrections to JT gravity with free matter are important for particles passing through the hatched region, even in the low energy (and large-$N$) limit. In the last section, we summarize our results and discuss some open questions.

\section{Coupled SYK dots at large $q$}
\label{sec:coupled-syks-at}

In this section, we describe the boundary model we study and the full solution for the two-point function. We will defer the majority of discussion of the bulk interpretation to the next sections. The key building block is the SYK ensemble of Hamiltonians on a system of $N$ Majorana fermions, the set of which we just denote by $\chi$ and elements $\chi_{j} \in \chi$ with $j \in 1, \ldots, N$; the elements obey the algebra $\{\chi_{j}, \chi_{k}\} = \delta_{jk}$. The ensemble is defined by an even number $q > 2$, a Gaussian random anti-symmetric tensor $J_{j_1 \cdots j_q}$, and Hamiltonian $H_{\text{SYK}}[J]$ with
\begin{align}
  \label{eq:syk-hamiltonian}
  H_{\text{SYK}}[\chi, J] &= \frac{i^{q/2}}{q!} \sum_{j_1, \ldots, j_q = 1}^N J_{j_1 \cdots j_q} \chi_{j_1} \cdots \chi_{j_q},\; \Delta = 1/q \\
  \langle J_{j_1 \cdots j_q} \rangle &= 0, \langle J_{j_1 \cdots j_q}^2 \rangle = J^2 \frac{(q-1)!}{N^{q-1}} = \mathcal{J}^2 \frac{2^{q-1}}{q} \frac{(q-1)!}{N^{q-1}}.
\end{align}

\subsection{Coupled SYK model and large-$N$ effective theory}
\label{sec:description-model}

Our model lives in a Hilbert space of $2N$ Majorana fermions. $N$ of the Majoranas are labeled ``left'' (``right''), and written $\chi_{Lj} \in \chi_L$ ($\chi_{Rj} \in \chi_R$). The total Hamiltonian we consider is then\footnote{The reason for the sign choice in $H^{L}$ will be explained below; here we note that $H^L$ and $H^R$ have the same spectrum since $H_{\text{SYK}}[\chi, J]^{T} = (-1)^{q/2} H_{\text{SYK}} [\chi^T, J]$ where $\chi^T$ means to transpose each Majorana operator in the set.}
\begin{align}
  H^L &= (-1)^{q/2} H_{\text{SYK}}[\chi_L, J], \; H^R = H_{\text{SYK}}[\chi_R, J] \\
  H(t) &= H^L(t) + H^R(t) + i \frac{\mu(t)}{q} \sum_{j=1}^N \chi_{Lj}(t) \chi_{Rj}(t)\label{eq:microscopic-coupled-syk-hamiltonian}
\end{align}
where $\mu(t)$ is some arbitrary non-negative function of time. Our goal is to understand correlation functions in a quantum quench problem, with the thermofield double state (described below) as the initial state, and time evolution determined by the time-dependent Hamiltonian (\ref{eq:microscopic-coupled-syk-hamiltonian}).

The thermofield double state is a special entangled state of two systems. For the left (right) fermions, this is a purification of the thermal density matrix for $H^L$ ($H^R$) at temperature $T = 1 / \beta$. We write the state $\ket{\text{TFD}}$, omitting a temperature label. If the eigenstates of $H^R$ are $\ket{E_n}$, then an explicit expression for the state is
\begin{align}
  \label{eq:tfd-bell-pair-state}
  \ket{\idop} &= \sum_n (\mathcal{T} \ket{E_n}) \ket{E_n} \\
  \label{eq:TFD-definition}
  \ket{\text{TFD}} &= \sum_n e^{- \beta E_n / 2} (\mathcal{T} \ket{E_n}) \ket{E_n} = e^{- \beta H^R / 2} \ket{\idop},
\end{align}
where the sum runs over energy eigenstates $\ket{E_n}$ and $\mathcal{T}$ is some anti-unitary transformation. The expression for $\ket{\idop}$ is unique up to an action of $O(D)$. The definition \eqref{eq:TFD-definition} is independent of the phase convention chosen for the $\ket{E_n}$, but \emph{does} depend on a choice of $\mathcal{T}$.

In our case, we determine the state $\ket{\idop}$ implicitly by a simple construction suited to the Hilbert space of the SYK model~\autocite{maldacena2018eternal}. Define $c_j = (\chi_{Lj} + i \chi_{Rj}) / \sqrt{2}$; then $\ket{\idop}$ is defined as the unique vacuum of the $c_j$. The sign in the definition of $H^L$ was chosen so that $(H^L - H^R)\ket{\idop} = (H^L - H^R)\ket{\text{TFD}} = 0$.

\begin{figure}[htb!]
  \centering
  \begin{subfigure}[b]{0.47\linewidth}
    \centering
    \begin{tikzpicture}[scale=4,
  state triangle/.style={fill, regular polygon, regular polygon sides=3, inner sep=1.1},
  boundary connector/.style={thin, densely dotted}]
  \def\contoureps{0.02}
  \def\reallength{0.7}
  \def\tickheight{0.05}
  \def\boundaryconnectorlength{0.1}
  \node[xslant=-0.3, state triangle, rotate=-30] (bra-1) at (0, 0) {};
  \fill[red] (bra-1.side 3) -- (bra-1.corner 2) -- (bra-1.corner 3) -- cycle;
  \fill[black] (bra-1.side 3) -- (bra-1.corner 1) -- (bra-1.corner 2) -- cycle;
  \draw[xslant=-0.3, boundary connector] (bra-1.corner 1) -- ++(0, \boundaryconnectorlength) coordinate (right contour start);
  \draw[contour, black]
  (right contour start) -- ++(1/4-\contoureps, 0)
  node[contour arrow, rotate=180] {}
  -- ++(0, \reallength)
  node[contour arrow, rotate=-90] {}
  -- ++(2*\contoureps, 0)
  -- ++(0, -\reallength)
  node[contour arrow, rotate=90] {}
  -- ++(1/4-\contoureps, 0)
  node[contour arrow, rotate=180] {}
  coordinate (right contour end);
  \draw[xslant=-0.3, boundary connector] (right contour end) -- ++(0, -\boundaryconnectorlength) coordinate (right ket start);
  \draw[xslant=-0.3, boundary connector, red]
  (bra-1.corner 3) -- ++(0, -\boundaryconnectorlength)
  coordinate (left contour start);
  \draw[contour, red] (left contour start)
  -- ++(1/4-\contoureps, 0)
  node[contour arrow, rotate=180] {}
  -- ++(0, \reallength)
  node[contour arrow, rotate=-90] {}
  -- ++(2*\contoureps, 0)
  -- ++(0, -\reallength)
  node[contour arrow, rotate=90] {}
  -- ++(1/4-\contoureps, 0)
  node[contour arrow, rotate=180] {}
  coordinate (left contour end);
  \draw[xslant=-0.3, boundary connector, red]
  (left contour end) -- ++(0, \boundaryconnectorlength);
  \node[xslant=-0.3,state triangle, rotate=30, anchor=corner 1] (ket-1) at (right ket start) {};
  \fill[red] (ket-1.corner 2) -- (ket-1.corner 3) -- (ket-1.side 1) -- cycle;
  \fill[black] (ket-1.corner 3) -- (ket-1.corner 1) -- (ket-1.side 1) -- cycle;
\end{tikzpicture}
    \label{fig:coherent-integration-schematic}
  \end{subfigure}
  \begin{subfigure}[b]{0.47\linewidth}
    \centering
    \begin{tikzpicture}[scale=4]
  \def\contoureps{0.02}
  \def\reallength{0.7}
  \def\tickheight{0.05}
  \draw[contour axes] (-1/2-\contoureps, 0) -- (1/2 + \contoureps, 0);
  \draw[contour tick] (-1/4, -\tickheight) -- (-1/4, \tickheight) node[above] {$- \beta / 4$};
  \draw[contour tick] (1/4, -\tickheight) node[below] {$\beta / 4$} -- (1/4, \tickheight);
  \draw[contour axes] (0, \reallength + \contoureps) -- (0, -\contoureps - \reallength);
  \draw[contour, red] (-1/2, 0) -- (-1/4 - \contoureps, 0) node[contour arrow] {}
  -- ++(0, -\reallength) node[contour arrow, rotate=-90] {}
  -- ++(2 * \contoureps, 0) -- ++(0, \reallength) node[contour arrow, rotate=90] {}
  -- (0, 0) node [contour arrow] {};
  \draw[contour, black] (0, 0)
  -- (1/4 - \contoureps, 0) node[contour arrow] {}
  -- ++(0, \reallength) node[contour arrow, rotate=90] {}
  -- ++(2 * \contoureps, 0) 
  -- ++(0, -\reallength) node[contour arrow, rotate=-90] {}
  -- (1/2, 0) node[contour arrow] {};
\end{tikzpicture}
    \label{fig:tfd-single-fermion-contour}
  \end{subfigure}
  \caption{(a) A schematic representation of the path integral construction from coherent states. Each triangle represents the infinite temperature thermofield double state $\ket{\idop}$, which enforces boundary conditions leading to \eqref{eq:single-chi-definition}, and the red and black lines represent the two SYK systems $L$ and $R$, respectively. The horizontal lines represent imaginary (Euclidean) time evolution while the vertical lines are real (Lorentzian) time evolution. (b) The path integral in (a) for two SYK systems is equivalent to a integration contour in a single copy of SYK model in the complex $\tau$ plane. The horizontal direction is $\Re\tau$ which is Euclidean time, and the vertical direction is $\Im\tau$, the Lorentzian time. The width of the vertical sections is infinitesimal, and the (minimum) height is set by the latest operator insertion. The red (black) lines denote parts of the integration corresponding to the left (right) fermions.}
  \label{fig:tfd-path-integration-contour}
\end{figure}
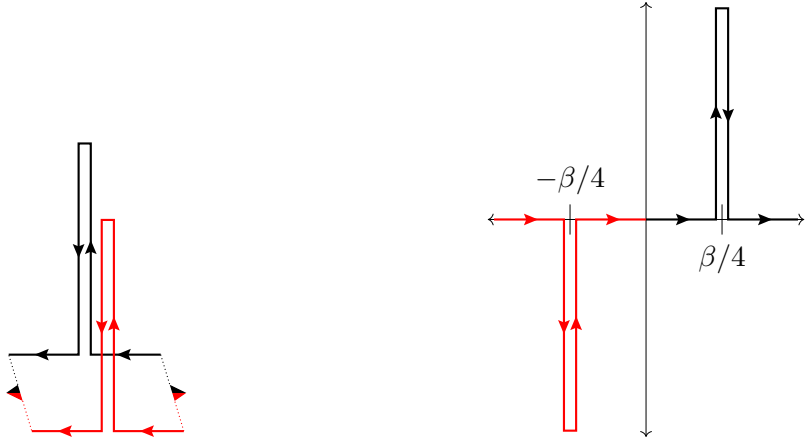

We are after correlation functions of the $\chi$ fermions, evolved by Hamiltonian $H$, in the thermofield double state $\ket{\text{TFD}}$. The simplest approach to finding the large-$N$ equations, which can be checked by a direct diagrammatic expansion, is through the path integral. We start by expanding correlation functions of the schematic form
\begin{equation}
  \bra{\text{TFD}} \chi_{Rj_n}(t_n) \chi_{Lj_{n-1}}(t_{n-1}) \cdots \chi_{R j_1}(t_1) \ket{\text{TFD}}
\end{equation}
in complex fermion coherent states. The path integral computes the correlation functions in Lorentzian time order. Due to the relations $H^R \ket{\idop} = H^L \ket{\idop}$ and $\chi_{L} \ket{\idop} = -i \chi_R \ket{\idop}$, it is possible to express correlation functions of the coupled model in a single copy language, as is illustrated in Fig. \ref{fig:tfd-path-integration-contour}. We can think of the state $\ket{\idop}$ as enforcing boundary conditions setting the Grassman fields $\chi_L = - i \chi_R$ at that point, whereas the state $\bra{\idop}$ enforces $\chi_L = i \chi_R$. Thus it is convenient to work in a complex time coordinate $\tau$ and a single anti-periodic (in $\tau \to \tau + \beta$) Grassman field
\begin{equation}
  \label{eq:single-chi-definition}
  \chi(\tau) =
  \begin{cases}
    \chi_R(\tau) & 0 \le \Re \tau \le \beta / 2 \\
    i \chi_L(- \tau) & - \beta / 2 < \Re \tau < 0
  \end{cases}, \; \chi(\beta / 4 + i t) = \chi_R(t), \, \chi(- \beta / 4 - i t) = i \chi_L(t)
\end{equation}
Then the thermofield double state is just constructed by imaginary time evolution with $H_{\text{SYK}}[\chi, J]$. The full action integral runs over the contour $C$ in $\tau$ shown in Figure~\ref{fig:tfd-path-integration-contour}. For convenience, the coupling is extended to the contour, with the constraint $\mu(\tau) = \mu(- \tau)$.
Then the full action is given by
\begin{align}
  S_{\text{SYK}} &= \int_C d\tau \frac{1}{2} \sum_{j = 1}^N \chi_j(\tau) \partial_{\tau} \chi_j(\tau) + H_{\text{SYK}}[\chi, J] \\
  S^{(0)}_{\mu} &= - \int_C d\tau \frac{\mu(\tau)}{2 q} \sgn (\Re \tau) \sum_{j=1}^N \chi_j(\tau) \chi_j(- \tau) \\
  S^{(0)} &= S_{\text{SYK}} + S^{(0)}_{\mu}.
\end{align}
We average over disorder, and introduce Hubbard Stratonovich fields $G$ and $\Sigma$. $\Sigma$ is a Lagrange multiplier setting
\begin{equation}
  G(\tau, \overline{\tau}) = \frac{1}{N} \sum_{j} \chi_j(\tau) \chi_j(\overline{\tau}).
\end{equation}
Replacing the fermions with $G$ in the SYK Hamiltonian and interaction, we are left with a quadratic action for $\chi$. Then we integrate out the fermions and find an action
\begin{align}
  S_{G\Sigma\text{SYK}} &= - \ln \Pf (\partial_{\tau} - \Sigma) + \frac{1}{2} \left( \int_C d\tau d\overline{\tau} \Sigma(\tau, \overline{\tau}) G(\tau, \overline{\tau}) - \frac{\mathcal{J}^2}{2 q^2} \left(2G(\tau, \overline{\tau})\right)^q \right) \\
  S_{G\Sigma \mu} &= - \int_C d\tau \frac{\mu(\tau)}{2 q} \sgn (\Re \tau) G(\tau, - \tau) \\
  \label{eq:g-sigma-interacting-action}
  S_{G\Sigma} &= N (S_{G\Sigma\text{SYK}} + S_{G \Sigma \mu}).
\end{align}
We emphasize that for $\mu$ not analytic, neither are $G$ or $\Sigma$. Thus this is just a compact way of writing equations for $G(\tau, \overline{\tau})$ and $\Sigma(\tau, \overline{\tau})$ fields not related by swaps of coordinates, where each coordinate is on various parts of the 6 contour segments. In the large $N$ limit, $G$ and $\Sigma$ are determined by the saddle point of action \eqref{eq:g-sigma-interacting-action}, which leads to the following Schwinger-Dyson equations:
\begin{align}
    G(\tau,\overline{\tau})&=\left(\partial_\tau -\Sigma\right)^{-1}\label{eq:Schwinger-Dyson 1}\\
    \Sigma(\tau,\overline{\tau})&=\frac{\mathcal{J}^2}{q}\left(2G(\tau,\overline{\tau})\right)^{q-1}+\frac{\mu(\tau)}q\sgn\left(\Re \tau\right)\delta\left(\tau+\overline{\tau}\right).\label{eq:Schwinger-Dyson 2}
\end{align} %
In the next subsection, we will solve the Schwinger-Dyson equation for generic $\mu(t)$ in the approximation $N \gg q^2 \gg 1$.

\subsection{Large-$q$ limit and the complex reparametrization dynamics}
\label{sec:solution-two-point}

In the large $q$ limit, it is convenient to change variables to $g$, defined by
\begin{align}
  \label{eq:G0-definition}
  G_0(\tau, \overline{\tau}) &= \frac{1}{2} \sgn (\tau - \overline{\tau}) \\
  \label{eq:G-largeq-form}
  G(\tau, \overline{\tau}) &= G_0(\tau, \overline{\tau}) \left(1 + \frac{g(\tau, \overline{\tau})}{q}\right) \approx G_0(\tau, \overline{\tau}) e^{g(\tau, \overline{\tau})/q}
\end{align}
where $\sgn (\tau - \overline{\tau})$ is $1$ if $\tau$ is later on the integration contour (with an end point at $- \beta / 2$) than $\overline{\tau}$, and $-1$ otherwise; $G_0=\partial_{\tau}^{-1}$ is the two-point function of a Majorana fermion with trivial dynamics. 
The self-energy $\Sigma$ is of order $O(1/q)$ when $\mathcal{J}$ and $\mu$ are kept constant in taking large $q$ limit. Taking a $\frac1q$ expansion of the Schwinger-Dyson equations one obtains\cite{maldacena2016remarks,eberlein2017quantum,maldacena2018eternal}
\begin{equation}
  \label{eq:large-q-concise-equation}
  - \partial_{\tau} \partial_{\overline{\tau}} g = 2 \mathcal{J}^2 e^g + \mu(\tau) \delta(\tau + \overline{\tau}) + O(1/q).
\end{equation}
We remind the reader that $g$ is defined on the contour $C$ in Figure~\ref{fig:tfd-path-integration-contour}. When either the $\tau$ or $\overline{\tau}$ coordinate is on a vertical segment of the contour, the right equations in real time are found by $\partial_{\tau} \to \pm i \partial_{t}$, where the sign is determined by the contour direction. When $\mu$ is not analytic, neither is $g$ so this is just a compact way of writing several independent equations.\footnote{Symmetries of our problem mean there are only 4 independent choices of segment pairings that determine the full two-point function on the whole contour; both coordinates on the real part of $C$, both on the right imaginary fold, one on the real part and one on the imaginary fold, and one on the left and one on the right imaginary fold. When both $\tau$ and $\overline{\tau}$ are real, from our initial conditions the solution $g(\tau, \overline{\tau})$ is the thermal correlator. We focus on just the Lorentzian time two-point functions for brevity (both coordinates having nonzero imaginary part).} The details of the method of solving this equation, subject to the boundary conditions imposed by the thermofield double state, are described in Appendix~\ref{sec:direct-deriv-hamilt}; here, we just explain the solution. (We note that there are other recent results on correlation functions in large-$q$ SYK model\cite{qi2019quantum,berkooz2019towards,choi2019exact,streicher2020syk} although they are not directly related to the goal of the current work.)

For simplicity, from now on we will adopt units such that
\begin{equation}
  \mathcal{J} = 1/2.
\end{equation}
We are interested in $g(\tau, \overline{\tau})$ on the Lorentzian time part of the contour $C$ (when $\tau,\overline{\tau}$ both have nonzero imaginary part). Without loss of generality, we will consider the region $\Re \tau > 0$, $\Re \tau \ge \Re \overline{\tau}$, and $\Im \tau \ge \Im \overline{\tau}$, which determines the other domains by symmetry. The real-time two-point functions are related to the contour two-point function $g(\tau,\overline{\tau})$ by the following equations:
\begin{align}
  G_R(u, u') &= 
  \frac{1}{N} \sum_{j=1}^N \langle \chi_{Rj}(u) \chi_{Rj}(u') \rangle
                                 = \frac{1}{2} e^{g_R(u, u')/q} \\
  i G_L(u, u') &= 
  \frac{i}{N} \sum_{j=1}^N \langle\chi_{Rj}(u) \chi_{Lj}(u')\rangle
                                   = \frac{1}{2} e^{g_L(u, u')/q},
\end{align}
with
\begin{align}
  g_{R}(u, u') &= g(\beta / 4 + iu, \beta / 4 + iu')
 \\
  g_{L}(u, u') &= g(\beta / 4 + iu, - \beta / 4 - iu').
\end{align}

It is useful to define particular solutions of a Loiuville equation,
\begin{align}
    g_{0R}(u) = - 2 \ln \sin \frac{u}{2} - i \pi  \label{eq:g0-definitions-right} \\
    g_{0L}(u) = - 2 \ln \cos \frac{u}{2}. \label{eq:g0-definitions-left}
\end{align}
We use the notation $S \in \{R, L\}$ to stand for either side, as in $g_{0S}$. The functions $g_{0S}$ solve the Liouville equations~\eqref{eq:large-q-concise-equation} away from the support of the $\delta$-function, with appropriate sign for each part of the contour. These $g_{0S}$ are not the solutions corresponding to the two-point function since they do not satisfy the correct boundary conditions. One condition is from the normalization of the fermions, $g(\tau, \tau) = 1$. Another condition comes from smoothness properties of $g$ and the term $\mu \delta(\tau + \overline{\tau})$ in the Liouville equation, given explicitly in \eqref{eq:non-operator-gtmt-bc} of Appendix~\ref{sec:direct-deriv-hamilt}. A third condition on $g$ is essentially the time derivative of the Hamiltonian, \eqref{eq:syken-derivative}, and a final condition \eqref{eq:syken-constraint} from $[\chi_{Rj}, [\chi_{Rj}, H]]$. Finally, we have initial conditions for $g$ from the thermofield double state.

As is well known~\autocite{liouville1853equation} (and proved in Appendix~\ref{sec:direct-deriv-hamilt}), all solutions to the ``bulk'' Liouville equations \eqref{eq:large-q-concise-equation} in our regions of interest can be (at least locally) written in the form (up to a shift by $i 2 \pi n$, $n \in \mathbb{Z}$)
\begin{equation}
  \label{eq:generic-liouville-reparametrization}
  g_S(u, u') = \ln \psi'_S(u) + \ln \overline{\psi}'_S(u') + g_{0S}(\psi_S(u) - \overline{\psi}_S(u'))
\end{equation}
for four independent complex functions $\psi_S$, $\overline{\psi}_S$, with $S=L$ or $R$.\footnote{The Liouville equation (away from any boundaries) is the equation of motion for a (1+1)-d CFT, and \eqref{eq:generic-liouville-reparametrization} can be thought of as a conformal transformation.}. The boundary conditions give differential equations for these functions. In Appendix~\ref{sec:direct-deriv-hamilt}, we show that these boundary conditions determine a unique two-point function, and in fact we can take
\begin{equation}
   \psi_R = \psi_L = \overline{\psi}_R^* = \overline{\psi}_L^*\equiv \psi.
\end{equation}
Thus the $g_S$ take the form
\begin{equation}
  \label{eq:g-reparametrization-transformation}
  g_{S}(u, u') = \ln \psi'(u) + \ln \psi'(u')^{*} + g_{0S}(\psi(u) - \psi(u')^{*}).
\end{equation}

Substituting the simpler ansatz \eqref{eq:g-reparametrization-transformation} into the boundary conditions, we obtain the equation of motions for $\psi(u)$. Defining $p$ to be the phase of $\psi'$ ($\psi' = |\psi'|e^{i p}$), we find the equations
\begin{align}
  |\psi'| = \sinh (|\Im \psi|), \; p' = - \mu + \frac{|\psi'|^2}{\sqrt{1 + |\psi'|^2}} \cos p;\label{eq:EOM psi}
\end{align}
more details are discussed in Appendix~\ref{sec:direct-deriv-hamilt}.

It turns out that these equations can be mapped to a system of Hamiltonian mechanics by a coordinate change. 
Define
\begin{equation}
  \label{eq:largeq-psi-phase-space-derivative}
  \psi'(u) = \frac{e^{ip}}{\sqrt{e^{2\phi}- 1}}
\end{equation}
and take $\phi$ and $p$ as a pair of canonical conjugate variables ($\{\phi, p\} = 1$). One can then verify that Eq. (\ref{eq:EOM psi}) is equivalent to the Hamiltonian equations of motion for the pair $(\phi,p)$ with the following Hamiltonian:
\begin{equation}
  \label{eq:largeq-effective-hamiltonian}
  \lgqeham = - \sqrt{1 - e^{- 2 \phi}} \cos p + \mu \phi.
\end{equation}
It is helpful to note that $\phi(u)$ has a simple physical interpretation. Using Eq. \eqref{eq:g-reparametrization-transformation} we obtain
\begin{equation}
    g_L(u, u) = - 2 \phi(u), \label{eq:gLphi}
\end{equation}
so that $\phi(u)$ directly determines the equal time two-point function between the two SYK islands. ($\phi(u)$ is always positive, which corresponds to $g_L(u,u)<0$ and $G_L(u,u)=e^{g_L(u,u)/q}<1$.) Larger $\phi$ corresponds to a weaker correlation between the two systems. 

The Hamiltonian \eqref{eq:largeq-effective-hamiltonian} is a central result in this work. This Hamiltonian determines the dynamics of the large-$q$ system (at least with the thermofield double initial state), and encodes the key differences between the finite energy dynamics (since the large-$q$ limit applies to all energy scales) and the more familiar low energy limit. Comparison of the large-$q$ and low energy limits will be discussed in the next section. Although we ``reverse-engineer'' $\lgqeham$ from the equation of motion, the Hamiltonian $\lgqeham$ also has a simple relation with the energy in the SYK system:
\begin{equation}
  \label{eq:largeqham-to-physical-energy}
  \langle H(t) \rangle = \frac{N}{q^2} \left( \lgqeham - \frac{\mu}{2 \Delta} \right)
\end{equation}
with $H(t)$ the coupled SYK Hamiltonian (\ref{eq:microscopic-coupled-syk-hamiltonian}). 
Useful consequences of the equations of motion are
\begin{equation}
  \Im \psi = - \tanh^{-1} e^{- \phi},\label{eq:Impsi}
\end{equation}
and
\begin{equation}
  \label{eq:largeq-schwarzian-like-energy}
  \frac{1}{2} \left( \frac{d \phi}{d u} \right)^2 + \frac{1}{2} \left[e^{- 2 \phi} +  (\mu \phi - \lgqeham)^2 - 1 \right] = 0.
\end{equation}
The initial conditions for $p$ and $\phi$ are set by their values in the thermofield double state:
\begin{align}
    \phi(0)&=\ln\csc \epsilon\label{eq:initial cond phi}\\
    p(0)&=0\label{eq:initial cond p}\\
    \text{with~}\sin \epsilon &= \frac{\pi - 2 \epsilon}{\beta \mathcal{J}},~0 < \epsilon < \frac{\pi} 2\label{eq:def epsilon}
\end{align}
To reiterate, to find the two-point function for arbitrary $\mu$ and TFD temperature, we integrate the equations of motion of $\phi(u),p(u)$ due to \eqref{eq:largeq-effective-hamiltonian}, and then find $\Re\psi$ by integrating \eqref{eq:largeq-psi-phase-space-derivative} with the initial condition $\Re\psi(0)=0$. ($\Im \psi(u)$ is directly determined by $\phi(u),p(u)$ due to Eq. \eqref{eq:Impsi}.)

As a trivial example, for $\mu=0$ we obtain the thermofield double solution (in Lorentzian time):
\begin{align}
  \phi(u) &= \ln \frac{\cosh (\sin (\epsilon) u)}{\sin (\epsilon)} \label{eq:largeq-tfd-phi} \\
  p(u) & = \tan^{-1} (\tan (\epsilon) \tanh (\sin (\epsilon) u)) \label{eq:largeq-tfd-p} \\
  \psi(u) &= 2 \tan^{-1} \tanh \frac{\sin(\epsilon) u - i \epsilon}{2} \label{eq:largeq-tfd-psi}.
\end{align}

\subsection{The constant $\mu$ solution}

As an example of our general results, in this subsection we study the constant $\mu$ system in some more details.

\begin{figure}
  \centering
  \tikzsetfigurename{largeqham-contours}
  \begin{tikzpicture}
  \begin{groupplot}[
    group style={group size=2 by 2,
      ylabels at=edge left,
      vertical sep={1.5cm}},
    ylabel shift={-1em},
    xlabel shift={-0.5em},
    width=7.7cm,
    contour/contour label style={font=\tiny},
    contour/label distance=6cm]
    \nextgroupplot[title={$\mu=0$}, largeq-phip-hamiltonian-3dplot-contour]
    \addphipHamContourPlot{0}
    \nextgroupplot[title={$\mu=10^{-2}$}, largeq-phip-hamiltonian-3dplot-contour,
    ylabel=\empty]
    \addphipHamContourPlot{1}
    \nextgroupplot[largeq-qp-hamiltonian-3dplot-contour]
    \addqpHamContourPlot{0}
    \nextgroupplot[largeq-qp-hamiltonian-3dplot-contour,
    ylabel=\empty]
    \addqpHamContourPlot{1}
    \fill[white, radius=0.5] (axis cs:0, 0) circle;
  \end{groupplot}
  \ifstandalone
  \else
  \node[below=1em,text width=1cm] at (group c1r1.south) {\subcaption{\label{fig:hphip-mu0}}};
  \node[below=1em,text width=1cm] at (group c2r1.south) {\subcaption{\label{fig:hphip-mu1}}};
  \node[below=1em,text width=1cm] at (group c1r2.south) {\subcaption{\label{fig:hqp-mu0}}};
  \node[below=1em,text width=1cm] at (group c2r2.south) {\subcaption{\label{fig:hqp-mu1}}};
  \fi
\end{tikzpicture}
  \caption{Plots of the Hamiltonian in the $\phi, p$ (\cref{fig:hphip-mu0,fig:hphip-mu1}) and $Q,P$ (\cref{fig:hqp-mu0,fig:hqp-mu1}) phase space coordinates for $\mu=0$ (\cref{fig:hphip-mu0,fig:hqp-mu0}) and $\mu=10^{-2}$ (\cref{fig:hphip-mu1,fig:hqp-mu1}). Contours followed by solutions are shown in white, and the contours at $-0.7$, 0, and 1 are highlighted in red color. In the $Q,P$ coordinate, $p$ is the angle with respect to the $Q$ axis, and $\sqrt{2 \phi}$ is the distance to the origin. For $\mu>0$ all orbits are closed and contain the fixed point at the minimum of $\lgqeham$. The origin in $Q,P$ coordinate is marked by a white dot in \cref{fig:hqp-mu1}, which lies on the $\lgqeham=0$ contour.} 
  \label{fig:H contour}
\end{figure}
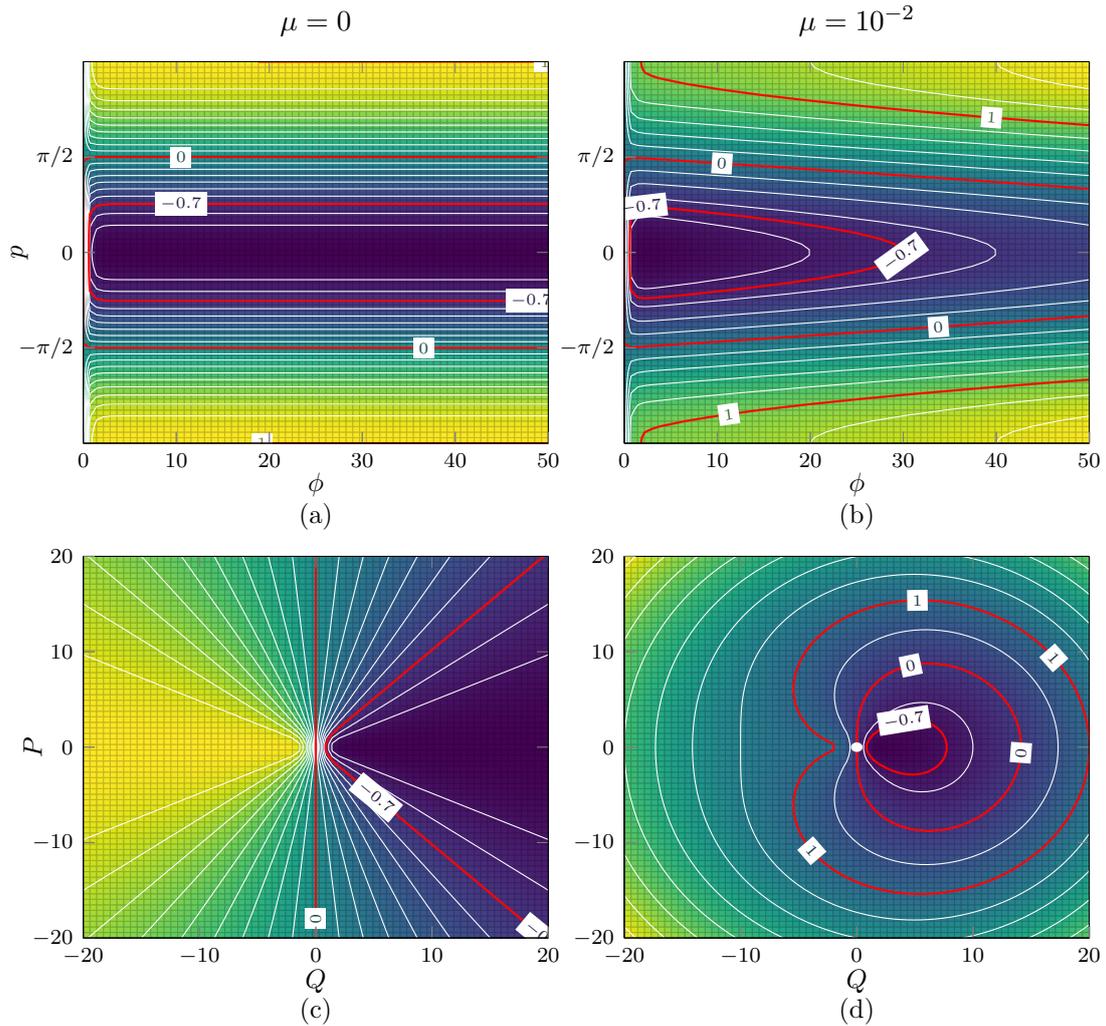

When $\mu$ is constant, the classical dynamics of $\phi,p$ conserves energy $H_Q$.
\Cref{fig:hphip-mu0,fig:hphip-mu1} show the energy contour plot of $H_Q$ versus $\phi$ and $p$ for $\mu=0$ and $\mu=10^{-2}$ respectively. For $\mu=0$, the orbits are non-compact, which simply corresponds to the thermofield double solution in Eq. (\ref{eq:largeq-tfd-phi})-(\ref{eq:largeq-tfd-psi}). $\phi$ approaches $+\infty$ in the past or future time infinity. For $\mu>0$, the orbits are compact in $\phi$ and periodic. Note from the perspective of the classical Hamiltonian system $p \sim p+2\pi$ should be thought of as an angle, and periodic in $2\pi$. For $\mu>0$ there are two kinds of orbits with different topology. For $\lgqeham<0$ the orbits have trivial winding of $p$, while for $\lgqeham>0$ the orbits have a nontrivial winding of $p$, such that $p$ changes by $2\pi$ during each period of motion. For visualizing the periodic $p$ variable it is helpful to consider a canonical transformation to a new pair of coordinates:
\begin{align}
  \label{eq:largeq-radial-canonical-transformation}
  Q &= \sqrt{2 \phi} \cos p, P = \sqrt{2 \phi} \sin p
  \end{align}
 The Hamiltonian is
  \begin{align}
  \lgqeham &= - \sqrt{1 - e^{- R^2}} \frac{Q}{R} + \frac{\mu}{2} R^2,~\text{with~}R \equiv \sqrt{P^2 + Q^2}=\sqrt{2\phi}
\end{align}
The equal energy contour in $Q,P$ plane are plotted in \cref{fig:hqp-mu0,fig:hqp-mu1}. $p$ is the angle coordinate in this plane. For $\mu>0$, the orbits are all closed in $Q,P$ plane, and the origin $(0,0)$ lies on the $\lgqeham=0$ contour. The $\lgqeham>0$ contours have a nontrivial winding around the origin. (The $Q,P$ coordinate is only used for illustration, and we will only use the $\phi,p$ coordinate for the rest of the paper.)

For each $\mu > 0$, there is always a minimum of $\lgqeham$, which occurs at $p = 0$ and $\phi = \phi_G(\mu)$. In terms of the coupling,
\begin{align}
  \phi_G(\mu) &= - \frac{1}{2} \ln \left( \mu \left[ \sqrt{\left( \frac{\mu}{2} \right)^2 + 1} - \frac{\mu}{2} \right] \right) \label{eq:minimum solution}
  \end{align}
  and the energy minimum is
  \begin{align}
  E_{QG}(\mu) &= - \sqrt{1 - e^{- 2 \phi_G(\mu)}} + \mu \phi_G(\mu).\label{eq:minimum energy}
\end{align}
For small $\mu$, $E_{QG}(\mu) \to -1$, while for large $\mu$, $E_{QG}(\mu) \approx - 1 / 2 \mu$. The minimal energy solution $\lgqeham=E_{QG}(\mu)$ is given by $\phi(u)=\phi_G(\mu),p(u)=0$, which corresponds to the following $\psi(u)$:
\begin{align}
  \label{eq:global-largeq-reparametrization}
  \psi_{G\mu}(u) &= V_G(\mu) (u - u_0) - i \tanh^{-1} e^{- \phi_G(\mu)} \\
  \label{eq:V_G-definition}
  V_G(\mu) &= \frac{e^{- \phi_G(\mu)}}{\sqrt{1 - e^{- 2 \phi_G(\mu)}}}
           = \sqrt{\mu \left( \frac{\mu}{2} + \sqrt{1 + \left( \frac{\mu}{2} \right)^2} \right)}.
\end{align}
Near the minimum, the Hamiltonian system is approximately harmonic with frequency
\begin{equation}
  \label{eq:largeq-hq-harmonic-frequency}
  \omega_G(\mu) = e^{- \phi_G(\mu)} \sqrt{\frac{3 + \coth \phi_G(\mu)}{2}}
  = \sqrt{2 \mu \sqrt{(\mu/2)^2 + 1}}.
\end{equation}
The static solution with energy $E_{QG}(\mu)$ corresponds has been studied in Ref.\cite{maldacena2018eternal}, which at low energy corresponds to the vacuum of a global AdS$_2$ spacetime, with the boundaries preserving the global time translation symmetry. More discussions about the bulk interpretation will be presented in Section~\ref{sec:bulk-interpretations}. In the following we will refer to this state as the ``global'' solution or ``fix point" solution.

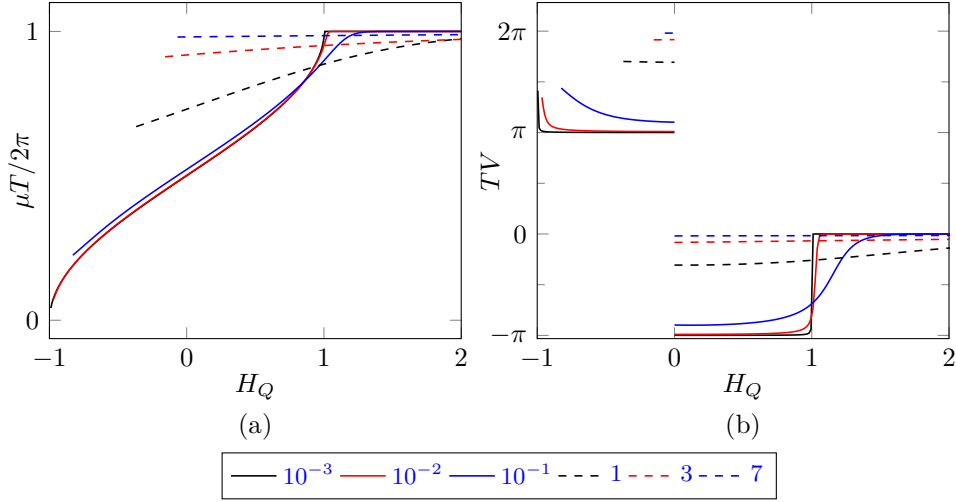
\begin{figure}[htb!]
  \centering
  \begin{tikzpicture}
  \pgfplotsset{
    every axis/.append style={
      no markers,
      cycle multi list={
        linestyles\nextlist
        black,red,blue
      }},
    every axis plot/.append style={
      semithick
    }}
  \begin{groupplot}[
    width=7cm,
    xlabel={$\lgqeham$},
    xmin={-1},
    xmax={2},
    xtick={-1,...,2},
    group style={group size=2 by 1, group name=vplots},
    ylabel shift={-1em},
    xlabel shift={-0.5em}]
    \nextgroupplot[ylabel={$\mu T / 2 \pi$},ymax={1.1},
    legend columns=-1,
    legend to name=klegend,
    ytick={0,1}]
    \addplot table[x=h,y=K0] {\plotdatadir/largeq-T.dat};
    \addlegendentry{$10^{-3}$}
    \addplot table[x=h,y=K1] {\plotdatadir/largeq-T.dat};
    \addlegendentry{$10^{-2}$}
    \addplot table[x=h,y=K2] {\plotdatadir/largeq-T.dat};
    \addlegendentry{$10^{-1}$}
    \addplot table[x=h,y=K3] {\plotdatadir/largeq-T.dat};
    \addlegendentry{$1$}
    \addplot table[x=h,y=K4] {\plotdatadir/largeq-T.dat};
    \addlegendentry{$3$}
    \addplot table[x=h,y=K5] {\plotdatadir/largeq-T.dat};
    \addlegendentry{$7$}
    \nextgroupplot[ylabel={$TV$},
    minor y tick num=1,
    ytick={-pi,0,pi,2*pi},
    yticklabels={$-\pi$, $0$, $\pi$, $2 \pi$},
    ymin={-pi-0.1}]
    \addplot table [x=h,y=K0] {\plotdatadir/largeq-tv.dat};
    \addplot table [x=h,y=K1] {\plotdatadir/largeq-tv.dat};
    \addplot table [x=h,y=K2] {\plotdatadir/largeq-tv.dat};
    \addplot table [x=h,y=K3] {\plotdatadir/largeq-tv.dat};
    \addplot table [x=h,y=K4] {\plotdatadir/largeq-tv.dat};
    \addplot table [x=h,y=K5] {\plotdatadir/largeq-tv.dat};
  \end{groupplot}
  \node[below=1.5em, text width=1cm] (cap1) at (vplots c1r1.south) {\subcaption{\label{fig:t-plot}}};
  \node[below=1.5em, text width=1cm] (cap2) at (vplots c2r1.south) {\subcaption{\label{fig:v-plot}}};
  \node at
  ($(cap1)!.5!(cap2) - (0,2em)$)
  {\ref{klegend}};
\end{tikzpicture}
  \caption{Plots of the period $T$ and velocity $V$ for various $\mu$ as a function of $\lgqeham$. For convenience we plot $T$ in units of $2\pi/\mu$ in \ref{fig:t-plot} and $V$ in units of $1/T$ in \ref{fig:v-plot}.}
  \label{fig:largeq-tv-plots}
\end{figure}

For more general orbits with energy $E>E_{QG}(\mu)$, $\phi(u)$ and $e^{ip(u)}$ are oscillating periodically. 
The period of an orbit can be obtained explicitly as the following integral\footnote{As a side remark, if we call the areas of the orbits in the $Q,P$ coordinate $A(\lgqeham)$ (see Fig. \ref{fig:H contour}), the period is simply given by $T(\lgqeham) = dA(\lgqeham)/d\lgqeham$.}:
\begin{equation}
  \label{eq:explicit-T-integral}
  T = 2 \int_{\phi_-}^{\phi_+} \frac{d \phi}{\sqrt{1 - e^{- 2 \phi} - (\mu \phi - \lgqeham)^2}};
\end{equation}
with $\phi_\pm$ ($\phi_- < \phi_+$) the two points on the orbit where $d\phi/du=0$. 
We plot the period as a function of $\lgqeham$ for various $\mu$ in Figure~\ref{fig:t-plot}. At low energy $\lgqeham\rightarrow E_{QG}(\mu)$, the period $T$ approaches the harmonic oscillation period $2\pi/\omega_G(\mu)$ with $\omega_G$ in Eq. \eqref{eq:largeq-hq-harmonic-frequency}. In the high energy limit $\lgqeham\gg 1$ the period approaches $2\pi/\mu$.

\begin{figure}[htb!]
  \centering
  \begin{tikzpicture}
  \newcommand{\repsitable}[1]{\plotdatadir/repsi-plots/mu#1.dat}
  \def\labels{$(1-10^{-4}) E_{QG}$,$(1-10^{-2})E_{QG}$,$(1-10^{-1}) E_{QG}$,$-10^{-2}$,$10^{-2}$,$10^{-1}$,$1$,$7$}
  \begin{groupplot}[
    group style={group size=2 by 1,
      ylabels at=edge left},
    ylabel={$\Re \psi / |VT|$},
    xlabel={$u/T$},
    ylabel shift={-0.5em},
    table/x=uot,
    every axis/.append style={
      no markers,
      cycle multi list={
        linestyles\nextlist
        black,red,blue
      }},
    every axis plot/.append style={
      semithick
    },
    width=7.7cm,
    legend columns=4,]
    \nextgroupplot[title={$\mu=10^{-2}$}, legend to name=klegend]
    \foreach \legendlabel [count=\ix from 0] in \labels {
      \addplot table[y=e\ix] {\repsitable{0}};
      \edef\temp{\noexpand\addlegendentry{\legendlabel}}\temp
    }
    \nextgroupplot[title={$\mu=1$}]
    \foreach \legendlabel [count=\ix from 0] in \labels {
      \addplot table[y=e\ix] {\repsitable{1}};
    }
  \end{groupplot}
  \node[below=1.5em, text width=1cm] (cap1) at (group c1r1.south) {\subcaption{\label{fig:small-k-psi-p}}};
  \node[below=1.5em, text width=1cm] (cap2) at (group c2r1.south) {\subcaption{\label{fig:large-k-psi-p}}};
  \node at
  ($(cap1)!.5!(cap2) - (0,2.5em)$)
  {\ref{klegend}};
\end{tikzpicture}
  \caption{Plots of $\Re \psi$ for small ($\mu = 10^{-2}$) and large ($\mu = 1$) coupling at various energies $\lgqeham$. Note that the net change of $\Re\psi$ over a period $T$ changes sign at zero energy.}
  \label{fig:repsi-plots}
\end{figure}
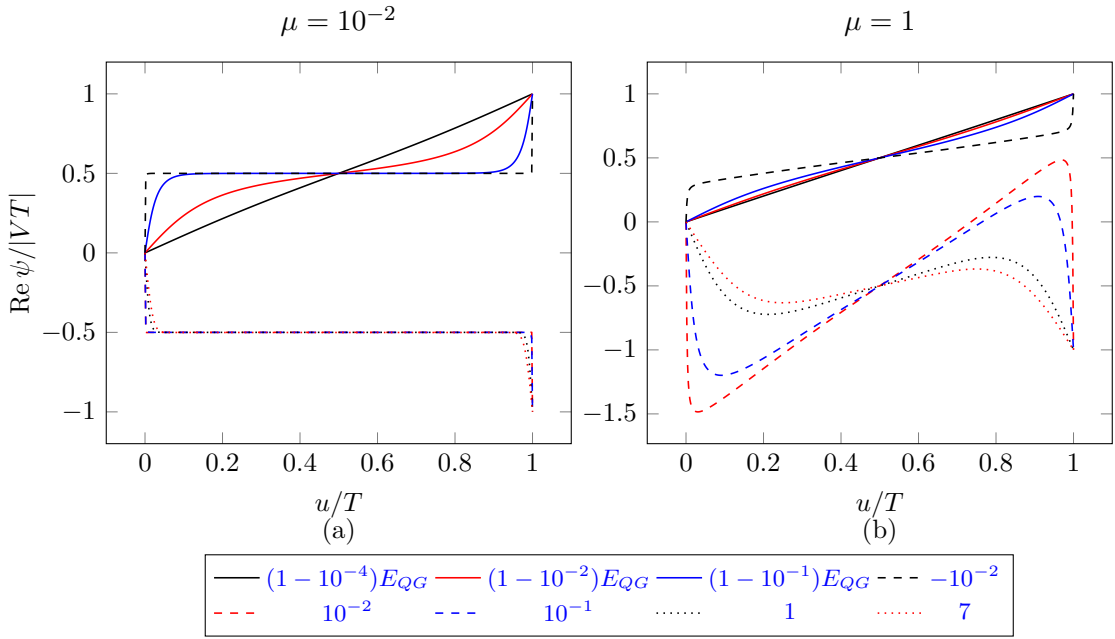

When $\phi$ and $e^{ip}$ are periodic, so is $\Im\psi(u)$ (see \eqref{eq:Impsi}), but $\Re\psi(u)$ is not periodic. Since $\Re\psi'(u)$ is periodic, $\psi(u)$ in general has the following form:
\begin{equation}
  \label{eq:largeq-psi-generic-V-form}
  \psi(u) = V(u - u_0) + \psi_P(u)  - i \tanh^{-1} e^{- \phi}
\end{equation}
where $V$ denotes the average velocity of $\Re\psi(u)$, which is a real constant depending on the value of $\lgqeham$ (and $\mu$). $u_0$ is an arbitrary real constant, and the second and third term on the right-hand side of the equation are both periodic with period $T$.\footnote{$\psi_P$ is odd over the period if $u$ is measured from one of the two points along the orbit that $d \phi / d u = 0$.} $V$ is given by the integral
\begin{equation}
  \label{eq:largeq-TV-integral}
  T V = \frac{1}{2} \oint \frac{\mu \phi - \lgqeham}{\sinh \phi} \mathrm{d} t
  = \int_{\phi_-}^{\phi_+} \frac{\mu \phi - \lgqeham}{\sinh \phi} \frac{\mathrm{d} \phi}{\sqrt{1 - e^{- 2 \phi} - (\mu \phi - \lgqeham)^2}}
\end{equation}
where the first integral is over a period of the orbit in phase space. It can be shown from \eqref{eq:largeq-TV-integral} that for $\lgqeham < 0$, $V > 0$, while for $\lgqeham > 0$, $V < 0$. The discontinuity at $\lgqeham = 0$ in $T V$ can be shown to be $2 \pi$. For large $\lgqeham$, $V \to 0^{-}$, since the motion becomes just winding around the $p$ cylinder at speed $\mu$. Some examples of the constant $V$ for various $\mu$ as a function of $\lgqeham$ are shown in Figure~\ref{fig:v-plot}. We show $\Re \psi$ for $\mu=10^{-2}$ in Figure~\ref{fig:small-k-psi-p}, and for larger $\mu = 1$ in Figure~\ref{fig:large-k-psi-p}.

In the remainder of the work we will focus on solutions with $\lgqeham<0$ since our main interest is in the comparison of large-$q$ theory and low energy theory. It is an interesting open question what is the physical interpretation of the different winding number of $p$ and the different sign of $V$ for $\lgqeham>0$ orbits.


\section{Comparison with low energy theory and the holographic dual interpretation}
\label{sec:bulk-interpretations}

To the leading order in the limit $N \gg \beta \mathcal{J} \gg 1$ (we call this the ``low energy'' limit), various aspects of the SYK model are described by a dual theory: $d=1+1$ Jackiw-Teitelboim dilaton (JT) gravity. The thermofield double state of the SYK model with decoupled dynamics corresponds to a two-sided black hole geometry, with no causal contact between the two boundaries. Maldacena and Qi~\autocite{maldacena2018eternal} proposed that the ground state of the coupled SYK model, with Hamiltonian \eqref{eq:microscopic-coupled-syk-hamiltonian} for small constant $\mu$, is dual to a global AdS$_2$ geometry with two boundaries in causal contact: an ``eternal traversable wormhole'' solution. They also studied low energy excitations in this model, including small fluctuation of the boundary, and conformal perturbations that correspond to bulk matter fields. 

In this section, we generalize the low energy discussion of the bulk dual theory to the large $q$ system with finite coupling and finite energy. We first review the bulk dual theory of the coupled SYK model for $\beta \mathcal{J} \gg 1$ and $\mu \ll \mathcal{J}$ proposed in Ref.~\autocite{maldacena2018eternal} (but for time dependent coupling $\mu(t)$), where $\beta$ is the temperature of the initial thermofield double state. Then we show how the large-$q$ effective theory correctly reproduces the low energy bulk theory in this limit. Beyond the low energy limit, we don't have a complete bulk dual theory, but it is helpful to still use the AdS$_2$ picture and view the theory as a 2d gravity with modified dynamics. We will discuss how time-dependent coupling $\mu(u)$ can be used to generate a generic solution that corresponds to a generic boundary location in AdS$_2$. 

\subsection{Bulk dual of the SYK model at low energy}
\label{sec:mapping-large-q}

To develop a bulk interpretation of our large-$q$ results, it is helpful to first review the duality in the low energy limit, based on Ref. \cite{maldacena2016conformal,maldacena2018eternal}. We start by describing the bulk theory, explain the correspondence to certain aspects of SYK physics, and list some properties that will later be compared to the large-$q$ theory.

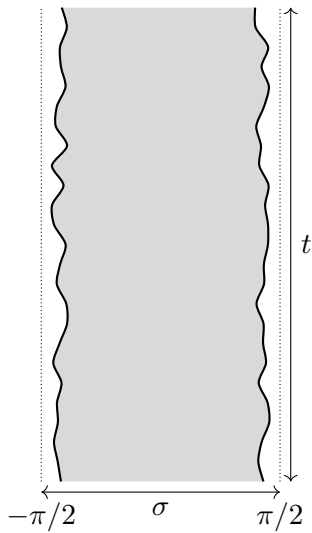
\begin{figure}[htb]
  \centering
  \begin{tikzpicture}
  \draw[conformal boundary]
  (-pi/2, -pi/2) coordinate (ads south west) -- (-pi/2, 3*pi/2)
  coordinate (ads north west) coordinate[midway] (ads west);
  \draw[conformal boundary]
  (pi/2, -pi/2) coordinate (ads south east) -- (pi/2, 3*pi/2)
  coordinate (ads north east);

  \pgfmathsetseed{7}
  \path[fill=gray!30] plot[smooth, domain=(-pi/2:3*pi/2)]
  ({rand*.1 + (pi/2 - 0.24)}, \x)
  -- plot[smooth, domain=(3*pi/2:-pi/2)]
  ({rand*.1 - (pi/2 - 0.24)}, \x)
  -- cycle;
  \pgfmathsetseed{7}
  \draw[syk boundary] plot[smooth, domain=(-pi/2:3*pi/2)]
  ({rand*.1 + (pi/2 - 0.24)}, \x);
  \draw[syk boundary] plot[smooth, domain=(3*pi/2:-pi/2)]
  ({rand*.1 - (pi/2 - 0.24)}, \x);

  \coordinate[below=4] (sigma axis start) at (ads south west);
  \coordinate[below=4] (sigma axis end) at (ads south east);
  \draw[<->] (sigma axis start) node[below] {$-\pi/2$}
  -- (sigma axis end) node[midway, below] {$\sigma$}
  node[below] {$\pi/2$};

  \coordinate[right=4] (t axis start) at (ads south east);
  \coordinate[right=4] (t axis end) at (ads north east);
  \draw[<->] (t axis start) -- (t axis end) node[midway,right] {$t$};
\end{tikzpicture}
  \caption{An example of the embedding of some generic solution of the JT action \eqref{eq:generic-jt-matter-action} in maximally extended $\ads_2$, in coordinates $(t,\sigma)$. 
  }
  \label{fig:ads2-coord-and-embedding}
\end{figure}

Consider then JT gravity with matter that is minimally coupled to the metric; the action is
\begin{equation}
  \label{eq:generic-jt-matter-action}
  \frac{1}{8 \pi G_N} \left[ \int_M \Phi \left( \frac{R}{2} + (2 \mathcal{J})^2 \right) + \int_{\partial M} \Phi k_G \right] + S_M
\end{equation}
where $S_M$ is the matter action, which we assume is local and does not involve the dilaton explicitly. Here, $\mathcal{J}^{-1}$ is some length scale, which we have chosen to be equal to the inverse coupling of the SYK model. Indeed, the equations of motion for $\Phi$ set the spacetime to be locally $\ads_2$ with curvature scale $(2 \mathcal{J})^{2}$. Globally, the spacetime has two timelike boundaries. Spacetimes satisfying the equations of motion can be isometrically embedded in $\ads_2$, so we can just imagine we are solving for the locations of the two boundaries in $\ads_2$. The solutions described this way will have an $\text{SL}(2; \mathbb{R})$ gauge symmetry arising from isometries of $\ads_2$, such that embeddings related by $\text{SL}(2; \mathbb{R})$ correspond to the same physical solution. Setting a boundary condition on $\Phi$ (which we will just take to be constant in this work) determines the relative locations of the boundaries, and an embedding in $\ads_2$ up to $\text{SL}(2; \mathbb{R})$. An illustration of this procedure, along with useful coordinates on $\ads_2$, is shown in Figure~\ref{fig:ads2-coord-and-embedding}. The metric in those coordinates is
\begin{equation}
  ds^2 = \frac{- dt^2 + d\sigma^2}{(2 \mathcal{J})^2 \cos^2 \sigma};
\end{equation}
$t$ and $\sigma$ are dimensionless. We will also use the dimensionful coordinates $\hat{t} = t / (2 \mathcal{J})$, $\hat{\sigma} = \sigma / (2 \mathcal{J})$. Occasionally, we will also use two other coordinates that cover part of AdS$_2$: the Poincare coordinate $ds^2=\frac{-dt_P^2+dz^2}{(2\mathcal{J})^2z^2}$, and the Rindler coordinate $ds^2=\left(2\mathcal{J}\right)^{-2}\left[-\sinh^2\rho dt_R^2+d\rho^2\right]$. The embeddings of three less generic solutions that are of particular interest to us are shown in Figure~\ref{fig:basic-ads2-embedded-geometries}.

To be concrete and to make comparisons with the SYK model simple, we set $\Phi|_{\partial M} = 1$, and
\begin{equation}
  G_N = \frac{q^2}{4 \pi N \hat{\alpha}_s}, \; \hat{\alpha}_S = 4 q^2 \alpha_S
\end{equation}
where $\alpha_S$ is a numerical constant found in \autocite{maldacena2016remarks} which approaches $1 / 4 q^2$ at large $q$. We imagine a scenario where $G_N \ll 1$ and $G_N \ll \Phi$. Then we can impose the equations of motion for $\Phi$, which leaves only the boundary term in the dilaton action. If we assume that the local velocity and acceleration of the boundary in (say) the $\sigma$ direction is small\footnote{Precisely, if $\phi$ is the geodesic coordinate parallel to $\partial_{\hat{\sigma}}$, the velocity (acceleration) we are referring to is $d \phi / d\hat{u}$ ($d^2 \phi / d\hat{u}^2$).}, dropping terms subleading in the velocity and acceleration we find an action
\begin{equation}
  S_g=\label{eq:dilaton-schwarzian-effective-action}
  - \frac{N \hat{\alpha}_S}{2 q^2} \int \frac{d \hat{u}}{2 \mathcal{J}} (\{ \tan (\mathcal{J} \hat{t}_R), \hat{u} \} + \{ \tan (\mathcal{J} \hat{t}_L), \hat{u} \} - 4 \mathcal{J}) + S_M
\end{equation}
where $\hat{t}_R$ ($\hat{t}_L$) is the global time coordinate of the embedding of the right (left) boundary and $\hat{u}$ is the proper time along either boundary. The notation $\{f, \hat{u}\}$ is the Schwarzian derivative of $f$ with respect to $\hat{u}$. Consistency requires that there is an embedding with $\hat{t}_R' \ll 1$, so we can use the condition that $\hat{u}$ is the proper time to approximately determine a particular embedding of the boundaries by
\begin{equation}
  \label{eq:low-energy-global-spatial-coordinate}
  \hat{\sigma}_{R/L}(\hat{u}) \approx \pm \left(\frac{\pi}{4 \mathcal{J}} - \hat{t}_{R/L}'(\hat{u})\right).
\end{equation}

Now, we consider the matter. After imposing the dilaton equations of motion, the matter action is a functional of the boundary locations. At fixed boundary location, it is just the action of matter fields propagating on an $\ads_2$ background with boundary. We imagine there are $N$ free fermion fields $\chi_j(x)$ with mass $m = \Delta - 1/2$, dual to boundary fields of dimension $\Delta = 1/q$, and call the fields at the left (right) boundary $\chi_{Lj}(\hat{u})$ ($\chi_{Rj}(\hat{u})$). (More precisely, the fermions can be interacting but their interaction should be suppressed by $\frac 1N$. For matter interaction in the dual theory of SYK model, see Ref. \cite{gross2017bulk}.) We normalize the fields so that at large geodesic distance in the Poincare coordinate at equal $z \ll 1$, their two-point function at $\hat{u} > \hat{u}'$ is
\begin{equation}
  \label{eq:bulk-matter-poincare-twopt}
  \langle \chi_{Rj}(\hat{u}) \chi_{Rj}(\hat{u}')\rangle = \frac{\hat{c}_{\Delta}}{2} \frac{e^{- i \pi \Delta}}{(\mathcal{J} (\hat{u} - \hat{u}'))^{2 \Delta}}, \;
  \hat{c}_{\Delta} = \left[ (1 - 2\Delta) \frac{\tan (\pi \Delta)}{\pi \Delta} \right]^{\Delta},
\end{equation}
and likewise for the $\chi_{Lj}$, with the same $\hat{c}_{\Delta}$. The normalization is chosen so that this two-point function is related to the low-energy SYK two-point functions by a reparametrization. The two-point function for points on boundaries satisfying the conditions above is given by a reparametrization of \eqref{eq:bulk-matter-poincare-twopt}. In particular, we have
\begin{equation}
  \label{eq:bulk-opposite-side-2pt-reparam}
  \langle \chi_{Rj} (\hat{u}) \chi_{Lj} (\hat{u}') \rangle
  = - i \frac{\hat{c}_{\Delta}}{2} \left( \frac{\hat{t}_R'(\hat{u}) \hat{t}_{L}'(\hat{u}')}{\cos^2 (\mathcal{J} (\hat{t}_R(\hat{u}) - \hat{t}_L(\hat{u}')))} \right)^{\Delta}.
\end{equation}

In the SYK model, at low energies there is an emergent conformal ({\it i.e.} reparametrization) symmetry which is weakly broken. The two-point function of Majorana fermions transforms by a reparametrization under this symmetry, and the reparametrizations get an action due to the explicit breaking of the symmetry. An effective action for these reparametrizations is believed to be exactly the dilaton part of \eqref{eq:dilaton-schwarzian-effective-action}. In both cases, we consider adding an interaction $i \frac{\hat{\mu}}{q} \sum \chi_{Lj} \chi_{Rj}$. In the gravitational model, we can imagine splitting the path integral into an outer one over the boundaries and an inner one over the matter. In the SYK model, we can imagine an outer integral over the soft reparametrization modes and an inner one over all other modes. In either case, call $\langle \cdot \rangle'$ the expectation taken in the inner integral, at some fixed $\hat{t}_R$ and $\hat{t}_L$. For small enough $\hat{\mu}$, we approximate in either case
\begin{multline}
  \langle e^{\int d\hat{u} \frac{\hat{\mu}(\hat{u})}{q} \sum_{j = 1}^{N} \chi_{Lj}(\hat{u}) \chi_{Rj}(\hat{u}) } \cdots \rangle'
  \approx \\ \exp \left(
    i \frac{N \hat{\alpha}_S}{2 q^2} \int d\hat{u} \frac{\mu(\hat{u})}{\Delta} \left( \frac{\hat{t}_R'(\hat{u}) \hat{t}_{L}'(\hat{u})}{\cos^2 (\mathcal{J} (\hat{t}_R(\hat{u}) - \hat{t}_L(\hat{u})))} \right)^{\Delta}
  \right) \langle \cdots \rangle'
\end{multline}
where $\mu = \hat{c}_{\Delta} \hat{\mu} / \hat{\alpha}_S$ and $\cdots$ stands for other operator insertions. We note here that if the bulk matter is free, this approximation is actually the exact result in the gravity theory.

Then the full effective low energy action for reparametrizations with the above approximations for both models in units $\mathcal{J} = 1/2$ (or times $u = 2 \mathcal{J} \hat{u}$, $t_S = 2 \mathcal{J} \hat{t}_S$) is
\begin{equation}
  S_S=\frac{N \hat{\alpha}_S}{2 q^2} \int du - (\{ \tan \frac{t_R}{2}, u \} + \{ \tan \frac{t_L}{2}, u \} - 2) + \frac{\mu}{\Delta} \left( \frac{t_R' t_L'}{\cos^2 \frac{t_R - t_L}{2}} \right)^{\Delta}.
\end{equation}
Consider the case that $t_L = t_R = t$\footnote{For solutions with no matter, ref. \autocite{maldacena2018eternal} show this can be arranged by a gauge transformation.}, and define $t' = e^{- \hat{\phi}}$. The action becomes
\begin{equation}
  \label{eq:linear-dilaton-free-matter-action}
  S_S=\frac{N \hat{\alpha}_S}{q^2} \int du \, \hat{\phi}'' + \frac{1}{2} (\hat{\phi}')^2 - \frac{1}{2} \left( e^{- 2 \hat{\phi}} - \frac{\mu}{\Delta} e^{- 2 \Delta \hat{\phi}} - 2 \right).
\end{equation}
We drop the total derivative $\hat{\phi}''$ term, which does not affect the local equations of motion.

For comparison with large-$q$ results, we note some features of the theory \eqref{eq:linear-dilaton-free-matter-action}. First, the Hamiltonian corresponding to this action is
\begin{equation}
  \label{eq:schwartzian-theory-hamiltonian}
  H_{S}=\frac{N \hat{\alpha}_S}{q^2} \left( \frac{1}{2} \hat{p}^2 + \frac{1}{2} \left( e^{- 2 \hat{\phi}} - \frac{\mu}{\Delta} e^{- 2 \Delta \hat{\phi}} - 2 \right) \right)
\end{equation}
where $\hat{p}$ is conjugate to $\hat{\phi}$. When $\mu = 0$, the solutions are the dual of the TFD at various temperatures $\beta$. To make expressions more uniform between low energy and large-$q$, we parameterize $\beta$ instead by
\begin{equation}
  \sin \hat{\epsilon} = \frac{\pi}{\beta \mathcal{J}}.
\end{equation}
Then the explicit solutions are
\begin{equation}
  \label{eq:low-energy-tfd-solutions}
  \hat{\phi}(u) = \ln \frac{\cosh (\sin(\hat{\epsilon}) u)}{\sin (\hat{\epsilon})} \implies
  t(u) = 2 \tan^{-1} \tanh \frac{\sin(\hat{\epsilon}) u}{2}.
\end{equation}
Comparing this equation with the large-$q$ equation \eqref{eq:largeq-tfd-phi} we see that $\hat{\phi}(u)$ in the low energy theory and $\phi(u)$ in the large-$q$ theory have the same behavior except for the slightly different coefficient $\sin\hat{\epsilon}$ versus $\sin\epsilon$.

\subsection{Comparison of large-$q$ and low energy effective theory}

To understand the bulk interpretation of our large-$q$ results, we first show how to derive the same low energy theory from the low energy limit of the large-$q$ effective theory. Before that, we define what we mean by a low-energy limit of large-$q$ solutions. Clearly, one aspect is $\beta\mathcal{J}\gg 1$ such that $\epsilon \ll 1$ ($\epsilon$ is defined in Eq. (\ref{eq:def epsilon})). The other two conditions are  $e^{- 2 \phi} \ll 1$ and $|p| \ll 1$. This is true in the initial thermofield double solution for small $\epsilon$, and therefore will be true for finite time after we turn on any coupling $\mu$. Qualitatively, we can see from the discussion in Section~\ref{sec:solution-two-point} that $e^{- 2 \phi} \ll 1$ only fails to hold at some points on a fixed $\mu > 0$ orbit for $\lgqeham$ near enough to $0$. On the other hand, $|p| \ll 1$ on the entire orbit requires $\lgqeham$ near $E_{QG}(\mu)$. We will see below that, after relating the large-$q$ solutions to the bulk, these last two assumptions are analogous to the ones leading to the Schwarzian theory \eqref{eq:dilaton-schwarzian-effective-action}; roughly speaking, $e^{- 2\phi} \ll 1$ corresponds to the boundaries being far apart, and $|p| \ll 1$ corresponds to a small relative boundary velocity.

First, it is straightforward to check the well-known fact that the $\epsilon \ll 1$ limit of the TFD initial conditions \cref{eq:largeq-tfd-phi,eq:largeq-tfd-p,eq:largeq-tfd-psi} matches the low energy TFD solutions \eqref{eq:low-energy-tfd-solutions}. We note here that in the low energy limit as defined above $\Re \psi' \approx e^{- \phi}$, so $\phi$ and $\hat{\phi}$ play a similar role, and should be thought of as analogous. More interestingly, we obtain a derivation of the Hamiltonian \eqref{eq:schwartzian-theory-hamiltonian} by taking the large-$q$ Hamiltonian $\lgqeham$ and expanding it in the low energy. The physical energy (which we need for the coefficient of the action) is \eqref{eq:largeqham-to-physical-energy}, so the Hamiltonian which gives rise to the correct action (at large $q$) is
\begin{align}
  H(t) &= \frac{N}{q^2} \left( \lgqeham - \frac{\mu}{2 \Delta} \right) \approx \frac{N}{q^2} \left( \frac{1}{2} p^2 + \frac{1}{2} (e^{- 2\phi} + 2 \mu \phi - \mu / \Delta - 2) \right) \\
            &\approx \frac{N}{q^2}
            \left( \frac{1}{2} p^2 + \frac{1}{2} \left(e^{- 2 \phi} - \frac{\mu}{\Delta} e^{- 2 \Delta \phi} - 2 \right) \right).
\end{align}
where in the first line we take the low-energy approximation, and in the second we use $\Delta \to 0$. This is exactly the Hamiltonian~\eqref{eq:schwartzian-theory-hamiltonian}. 

From the derivation of the low energy theory, we see that at low energy it is natural to consider $\psi$ to be related to $t_R$, and $\psi^{*}$ to $t_L$. Indeed, in addition to the similarity in their low energy dynamics, both functions actually play the exact same role in determining the two-point function as $\Delta \to 0$: the two-point function is obtained by a reparametrization of the single- and two-sided global functions by $t_R$ and $t_L$, and from \cref{eq:G-largeq-form,eq:g0-definitions-right,eq:g0-definitions-left,eq:g-reparametrization-transformation} we see the same is true with $t_R \to \psi$ and $t_L \to \psi^*$ in the large-$q$ case. Concretely, we propose to understand the low energy dynamics and corrections to them by the analogy
\begin{equation}
  t_R = t_L = t \leftrightarrow t_G(u) = \Re \psi(u).
\end{equation}

To give the full embedding at low energy, we can find $\sigma$ by \eqref{eq:low-energy-global-spatial-coordinate} in the low energy limit. We can give some physical interpretations to $p$ and $\phi$ that help to understand the low energy limit. First, we notice that (at least for $e^{- 2 \phi} \ll 1$) it is consistent to take $2 \phi = - g_L(u, u)$ the geodesic distance between points on the boundary at the same $u$. This also gives $p \approx \sin p = \frac{d}{du} (\cosh^{-1} e^{\phi}) \approx d \phi / du$ the interpretation of the local velocity of the boundary in the $\sigma$ direction. Then $|p| \ll 1$ is consistent with the assumptions on the bulk that give rise to the Schwarzian action for the boundary. With these interpretations, one can easily see some basic features of the geometry from orbits of solutions in the $Q,P$ canonical coordinates (see for example \cref{fig:hqp-mu0,fig:hqp-mu1}).

To get some practice, we can consider the geometries corresponding to the two simplest cases, the decoupled case $\mu = 0$ and the $\mu > 0$ ground state case $\lgqeham = E_{QG}$. As a further simple consistency check, we will show that the two-point functions derived from large-$q$ are just regulated versions of the appropriate bulk correlator corresponding to the same geometry (c.f. \eqref{eq:bulk-opposite-side-2pt-reparam}). The $\mu = 0$ and a portion of the $\mu > 0$ fixed point geometry are shown in \cref{fig:rindler-sol-ads-embedding,fig:global-sol-ads-embedding}. In the $\mu = 0$ case, as we discussed in the $\epsilon \ll 1$ limit the boundaries approach boundaries at constant Rindler radius $\rho \approx \pm \ln \csc \epsilon$ under our mapping. The two-point functions are
\begin{equation}
  G_R(u, u') = \frac{e^{- i \pi \Delta}}{2} \left( \frac{\sin \epsilon}{\sinh \left(\frac{u - u'}{2} - i \epsilon\right)} \right)^{2 \Delta}, \;
  G_L(u, u') = - \frac{i}{2} \left( \frac{\sin \epsilon}{\cosh \left( \frac{u + u'}{2} \right)} \right)^{2 \Delta}
\end{equation}

The other simple case is when $\lgqeham = E_{QG}$. The full reparametrizations are \eqref{eq:global-largeq-reparametrization}. We have $t_G(u) = V_G(u - u_0)$ for some $u_0$, so the boundaries reside at constant global spatial coordinate $\sigma \approx \pi / 2 - e^{- \phi_G}$. The two-point functions are
\begin{align}
  G_R(u, u') &= \frac{e^{- i \pi \Delta}}{2} \left( \frac{V_G}{\sin(\frac{V_G(u - u')}{2} - i \tanh^{-1} e^{- \phi_G})} \right)^{2\Delta} \\
  G_L(u, u') &= - \frac{i}{2} \left( \frac{V_G}{\cos(\frac{V_G(u - u')}{2} - i \tanh^{-1} e^{- \phi_G})} \right)^{2\Delta}
\end{align}
which are exactly regulated versions of the bulk functions for boundaries at constant global $\sigma$ separation (c.f. \eqref{eq:bulk-opposite-side-2pt-reparam}). 

More generally, in either the low energy or large-$q$ theory, we can ``engineer'' a desired boundary by fixing the embedding, then using the equations of motion to find the required coupling. In the low energy case, this is
\begin{equation}
  \mu = e^{2 \Delta \hat{\phi}} (e^{- 2 \hat{\phi}} - \hat{\phi}'') \approx e^{- 2 \hat{\phi}} - \hat{\phi}''.
\end{equation}
In the large $q$ theory we instead have 
\begin{equation}
  \mu = \frac{\sec p}{\sqrt{1 - e^{- 2 \phi}}} (e^{- 2 \phi} - \phi'').
\end{equation}
As expected, in the low energy limit the two equations above agree with each other. We will discuss a particular example of the rescued black hole geometry in Section~\ref{sec:rescue}.

We briefly comment on what happens if we keep the geometrical interpretation in regions that are far from low energy as we have defined it. First, when we violate $e^{- 2 \phi} \ll 1$, then according to our mapping the boundaries begin to approach each other. At the same time, $\Re \psi'$ begins to blow up. This happens as $\lgqeham \to 0$. The differences are more dramatic when we violate $|p| \ll 1$. In particular, when $\lgqeham > 0$, we have regions where $\Re \psi' < 0$ (this is the $Q < 0$ half-plane or $|p| > \pi / 2$), so it is as though, under our mapping, bulk time runs backward. Interestingly, even if a majority of time in an orbit is spent with $\Re \psi' > 0$, once there is a region with $\Re \psi' < 0$ (i.e. $\lgqeham < 0$), $\Re \psi$ will always \emph{decrease} over the course of an orbit (this is just the statement about \eqref{eq:largeq-psi-generic-V-form} that $V < 0$). The detailed behaviour of $t_G$ on various orbits can be seen for small and large coupling and a variety of energies in \cref{fig:small-k-psi-p,fig:large-k-psi-p}.

\section{Correlation functions}
\label{sec:four-point-functions}

\subsection{The two-point function and spectrum}
\label{sec:two-point-function}

To highlight some of the corrections to the low-energy theory, we can study the two-point function, or equivalently $g_S(u_2, u_1)$, at fixed $\mu$. The Fourier transform of the two-point function in $u_2$ contains information about the excitation spectrum of the coupled SYK model. For example, the two-point function between two opposite boundaries is given by
\begin{align}
    G_{L}(u_2,u_1)&\simeq -\frac i2e^{g_L(u_2,u_1)/q}=-\frac i2\left(\frac{2\psi'(u_2)\psi'(u_1)^*}{\cos\left(\psi(u_2)-\psi(u_1)^*\right)+1}\right)^{1/q}\label{eq:twopt global}
\end{align}
where $\psi(u)$ is given by Eq. (\ref{eq:largeq-psi-generic-V-form}). As a reminder, $\psi(u)$ has the form $\psi(u)=Vu+\delta \psi(u)$, which is a sum of a linear $u$ term with slope $V$ and other terms $\delta\psi(u)$ that are all periodic in $u$ with period $T$ (given by Eq. \eqref{eq:explicit-T-integral}). Therefore the two-point function (\ref{eq:twopt global}) is a function of the term $e^{iV(u_2-u_1)}$ and other terms $\psi'(u)$,$\delta\psi(u)$ which all have period $T$. If we fix $u_1$ and take Fourier transform of the two-point function over $u_2$, in the $\lgqeham<0$ case we obtain a comb of $\delta$-functions at frequencies
\begin{align}
    \omega_{n_M,n_\gamma}\equiv (\Delta + n_M) V + n_{\gamma} \omega_{\gamma},\text{~with~}\omega_\gamma=\frac{2\pi}T\label{eq:all frequencies}
\end{align} 
for integers $n_{\gamma}$, $n_{M}$. As a reminder, $V$ and $T$ are determined by Eq. (\ref{eq:explicit-T-integral}) and  (\ref{eq:largeq-TV-integral}). We note that for $\lgqeham>0$ solutions, the frequency $V$ is replaced by $V+2\pi$ in the equation above, due to the nontrivial winding of $p$. In the following we will only discuss the $\lgqeham<0$ case.

When the energy is near the minimum ($H_Q\rightarrow E_{QG}(\mu)$), our result is a finite energy generalization of the ``traversable wormhole'' solution in ref. \autocite{maldacena2018eternal}, and we can compare the frequencies with the results there. In this limit we have
\begin{align}
V = V_G(\mu) = t_G' = \sqrt{\mu \left( \frac{\mu}{2} + \sqrt{1 + \left( \frac{\mu}{2} \right)^2} \right)} \\
\omega_{\gamma} \to \omega_{G} = \sqrt{2 \mu \sqrt{1 + \left( \frac{\mu}{2} \right)^2}}.
\end{align}
where we recall that $\Re \psi' = t_G' = V_G$ at the fixed point (see \eqref{eq:global-largeq-reparametrization}), and $\omega_G$ is just the harmonic frequency at the minimum of $\lgqeham$, \eqref{eq:largeq-hq-harmonic-frequency}.

For comparison, the low energy theory gives an excitation spectrum with ``matter'' excitations having a gap $\Delta t'$ and energy spacing $t'$, with $t' = \mu^{1 / 2(1-\Delta)}$, as well as ``gravitational'' excitations with energy spacing $\hat{\omega}_{\gamma} = t' \sqrt{2(1 - \Delta)}$. This agrees with our results in $\mu\ll 1$ and $\Delta=1/q\ll 1$ limit, with $t'$ corresponding to $V$ and $\hat{\omega}_\gamma$ corresponding to $\omega_\gamma$. From this comparison we see that the two unit frequencies $V$ and $\omega_\gamma$ in Eq. (\ref{eq:all frequencies}) can be interpreted as the energy unit of matter field excitation and that of gravitational excitation, respectively. 

In particular, the ratio of these two frequencies in the large-$q$ theory is given by
\begin{equation}
  \frac{V}{\omega_\gamma} = \frac{1}{\sqrt{2}} \sqrt{1 + \frac{\mu/2}{\sqrt{1 + (\mu/2)^2}}},\label{eq:frequency ratio}
\end{equation}
which approaches $1/\sqrt{2}$ in low energy limit, but gets a nontrivial correction for finite $\mu$. Note that exactly {\it at} the fixed point (minimal energy solution), we will only have a single series of frequencies $(\Delta + n_M) V$, because the boundary location is time translation invariant. In this case the ``gravitational'' frequency $\omega_\gamma$ still appears in four-point functions, as will be discussed in Section~\ref{sec:scatt-resc-geom}. 

\begin{figure}[htb!]
  \centering
  \begin{tikzpicture}
  \pgfplotsset{
    every axis/.append style={
      no markers,
      cycle multi list={
        linestyles\nextlist
        black,red,blue
      }},
    every axis plot/.append style={
      semithick
    }}
  \begin{axis}[xlabel={$\lgqeham$},
    ylabel={$\omega_{\gamma} / \omega_{M}$},
    xmin={-1.1},
    xmax=4,
    xtick={-1,...,4},
    ytick={1, sqrt(2), 2},
    yticklabels={$1$, $\sqrt{2}$, $2$},
    ylabel shift={-0.5em}]
    \foreach \kn in {0,...,7} {
      \addplot table [x=h,y expr= (2 * pi) / \thisrow{v2kT}] {\plotdatadir/psi-a-v2/psi-a-v2-K\kn.dat};
    }
    \foreach \klabel in {$10^{-3}$,$10^{-2}$,$10^{-1}$,$1/5$,$1/2$,$1$,$2$,$3$} {
      \edef\temp{\noexpand\addlegendentry{\klabel};}
      \temp
    }
  \end{axis}
\end{tikzpicture}
  \caption{The ratio of the two base frequencies that appear in the constant-$\mu$ two-point function as a function of energy $\lgqeham$ for various couplings. In the text, we focus on the case $\lgqeham < 0$, where $\omega_M = V$. When $\lgqeham > 0$, we have $\omega_M = V + 2 \pi$ (the formula \eqref{eq:all frequencies} holds at all energies if we make the replacement $V \to \omega_M$).}
  \label{fig:tkv2-plots}
\end{figure}
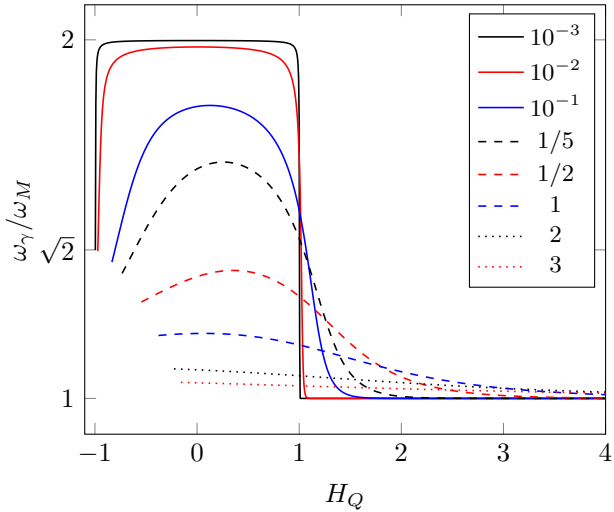

Beyond the limit $\lgqeham{}\rightarrow E_{QG}(\mu)$, the ratio of the two base frequencies depends on both $\mu$ and the energy of the solution, and is shown in Figure~\ref{fig:tkv2-plots} for a variety of $\mu$ and $\lgqeham$. If the ratio is rational, the set of frequencies $\omega_{n_M,n_\gamma}$ in Eq. (\ref{eq:all frequencies}) will be discrete. Otherwise it will be dense in $\mathbb{R}$.

\subsection{Four-point functions from response theory}
\label{sec:four-point-functions-1}

The complex reparametrization dynamics with Hamiltonian (\ref{eq:largeq-effective-hamiltonian}) determines not only the two-point functions but also certain types of higher point functions. Since we can obtain two-point functions for generic time-dependent coupling $\mu(u)$, we can vary $\mu(u)$ and study the change of two-point functions it induces. The first derivative of the fermion two-point function over $\mu(u)$ corresponds to a fermion four-point function. Higher derivatives can also be studied, but in this paper we will focus on the four-point function.

Consider the perturbation $\mu_{\nu}(u) = \mu(u) + \nu \delta(u - u_0)$. The response of two-point function $g_S(u_2,u_1)$ to such a perturbation with infinitesimal $\nu$ is the following four-point function:
\begin{align}
  \left. \frac{d g_{S}(u_2, u_1)}{d \nu} \right|_{\nu = 0}
  &=
    \begin{split}
    - \frac{2}{N} \sum_{j,k} \Big(\langle \mathcal{T} [ \chi_{Rj}(u_2) &\chi_{Sj}(u_1) \chi_{Rk}(u_0) \chi_{Lk}(u_0) ]\rangle - \\
    &\langle \chi_{Rk}(u_0) \chi_{Lk}(u_0) \mathcal{T} [\chi_{Rj}(u_2) \chi_{Sj}(u_1)]\rangle\Big)
  \end{split} \\
  &\equiv - \mathcal{F}_{RS}(u_2, u_1, u_0)
\end{align}
where we have explicitly indicated which operators are contour-ordered by the symbol $\mathcal{T}$, and $S \in \{R, L\}$ indicates the side of the fermion insertion at $u_1$. 

Since the two-point function is determined by the complex reparametrization $\psi(u)$, the four-point function above is determined by the equation of motion of $\psi(u)$, or equivalently that of the canonical variables $p,\phi$ defined in Eq. (\ref{eq:largeq-psi-phase-space-derivative}). The $\delta$-function shift of $\mu(u)$ leads to a jump in momentum $p\left(u_0^+\right)= p\left(u_0\right)-\nu$, since the canonical equation of motion $\dot{p}=-\partial H_Q/\partial \phi$ contains a term $-\mu$. Therefore the four-point functions are obtained by the derivative of the two-point function in Eq. (\ref{eq:g-reparametrization-transformation}) over $p(u_0)$, which has the following explicit expression:
\begin{align}
  \mathcal{F}_{RS}(u_2, u_1, u_0)
  &= (A(u_2, u_0) + A(u_1, u_0)^{*})
                          + \left( B(u_2, u_0) - B(u_1, u_0)^{*} \right) F_S(u_2,u_1) \\
  F_L(u_2, u_1)
  &= \tan \frac{\psi(u_2) - \psi(u_1)^{*}}{2} = \tan \left(  \frac{t_G(u_2) - t_G(u_1)}{2} - \frac{i}{2} \gamma(u_2, u_1) \right) \\
  F_R &= - 1 / F_L \\
  \label{eq:gamma-regulator-definition}
  \gamma(u_2, u_1) &= \tanh^{-1} \left(  \frac{e^{- \phi(u_1)} + e^{- \phi(u_2)}}{1 + e^{- (\phi(u_2) + \phi(u_1))}} \right),
\end{align}
where 
\begin{equation}
  A(u, u_0) \equiv \frac{d \ln \psi'(u)}{dp(u_0)}, \; B(u, u_0) \equiv \frac{d \psi(u)}{dp(u_0)}.\label{eq:defAB}
\end{equation}
Since $\psi'$ is a function of $\phi$ and $p$ (Eq. (\ref{eq:largeq-psi-phase-space-derivative})), the coefficients $A(u,u_0)$ and $B(u,u_0)$ are determined by the Poisson brackets $\{\phi(u_0), \phi(u)\}=d\phi(u)/dp(u_0)$ and $\{\phi(u_0), p(u)\}=dp(u)/dp(u_0)$. (The calculation of $B(u,u_0)$ requires an integration over time.) The Poisson brackets can be determined by solving a differential equation determined by the Hamiltonian $\lgqeham$. 

Details of the four-point function calculation are discussed in Appendix \ref{sec:details-four-point}. Here we will only present a simple example, to help give some physical intuition. For $u_1=u_2=u$, the four-point function $\mathcal{F}_{RL}(u,u,u_0)$ is simply a derivative of $g_L(u,u)=-2\phi$ (Eq. (\ref{eq:gLphi})), so that
\begin{equation}
 \mathcal{F}_{RL}(u,u,u_0)=-2\frac{d\phi(u)}{dp(u_0)}=2(1 - e^{- 2 \phi(u)}) \Re A(u, u_0).\label{eq:FRLuuu0}
\end{equation}
If we consider the fixed point (global) solution $\phi(u)=\phi_G(\mu)$ at some coupling $\mu > 0$ (with both times $u, u_0$ in this region), we obtain
\begin{equation}
  \mathcal{F}_{RL}(u, u, u_0) = - 4 \frac{\sinh(\phi_G(\mu))}{\sqrt{2 - e^{- 2 \phi_G(\mu)}}} \sin \left(\omega_G(\mu)\left(u-u_0\right)\right)\label{eq:FRLuuu0 global}
\end{equation}
As expected, the four-point function oscillates with $\omega_G$, the harmonic frequency at the minimum of $\lgqeham$ given in \eqref{eq:largeq-hq-harmonic-frequency}, which corresponds to a small oscillation of $\phi$ around the potential minimum $\phi_G$ (and $p$ about 0). The frequency $V$ that appears in the two-point function of the global solution does not appear in this four-point function. The amplitude grows exponentially with $\phi_G$. In the small coupling limit $\mu\ll 1$, we have $\omega_G\simeq\sqrt{2\mu}$ and the oscillation amplitude is approximately $\sqrt{2}e^{\phi_G(\mu)}\simeq \sqrt{2/\mu}$. 

\section{Rescued black hole geometry}
\label{sec:rescue}

\subsection{Rescuing a black hole by time-dependent coupling}
\label{sec:rescued-geometry}

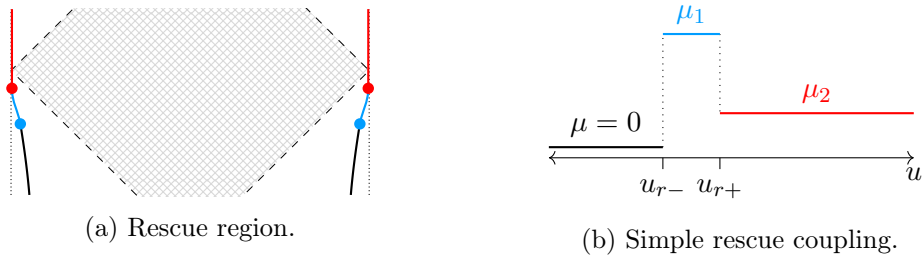
\begin{figure}
  \centering
  \begin{subfigure}{0.47\linewidth}
    \centering
    \begin{tikzpicture}[scale=1.5,
  mu1 style/.style={mu1color}]
  \tikzset{declare function={
      rightrindlertG(\tG,\rho) = acos(cos(\tG r) / cosh(\rho)) * pi / 180;
      leftrindlertG(\tG,\rho) = - rightrindlertG(\tG, \rho);
    }}
  \def\boundrho{2.4}
  \def\globallength{0.7}
  \pgfmathsetmacro{\lowercut}{pi/2*0.3}
  \pgfmathsetmacro{\muostart}{pi/2*0.7}
  \pgfmathsetmacro{\muoend}{pi/2*0.9}
  \pgfmathsetmacro{\globalsigma}{rightrindlertG(\muostart, \boundrho)*(1-0.9) + pi/2*0.9}
  \coordinate (global boundary start) at (\globalsigma, \muoend);
  \path (global boundary start) -- ++(0, \globallength) coordinate (global boundary end);

  \draw[conformal boundary] let \p1=(global boundary end) in
  (pi/2, \lowercut) -- (pi/2, \y1);
  \draw[conformal boundary] let \p1=(global boundary end) in
  (-pi/2, \lowercut) -- (-pi/2, \y1);

  \begin{scope}[overlay]
    \path[horizon region] (0, 0) -- (pi/2, pi/2) -- (0, pi) -- (- pi/2, pi/2) -- cycle;
    \draw[light ray marker] (0, 0) -- (pi/2, pi/2) -- (0, pi) -- (- pi/2, pi/2) -- cycle;
  \end{scope}

  \draw[syk boundary] plot[smooth,
  domain=(\lowercut):(\muostart)]
  ({rightrindlertG(\x, \boundrho)}, \x)
  coordinate (decoupled boundary end);
  \draw[syk boundary, red] (global boundary start) -- (global boundary end);
  \draw[syk boundary, mu1 style] (decoupled boundary end) to[out=80, in=-90] (global boundary start);

  \fill[boundary insertion point, mu1color] (decoupled boundary end) circle;
  \fill[boundary insertion point, red] (global boundary start) circle;
  \draw[syk boundary] plot[smooth,
  domain=(\lowercut):(\muostart)]
  ({leftrindlertG(\x, \boundrho)}, \x)
  coordinate (left decoupled boundary end);
  \coordinate (left global boundary start) at (- \globalsigma, \muoend);
  \draw[syk boundary, red] (left global boundary start) -- ++(0, \globallength) coordinate (left global boundary end);
  \draw[syk boundary, mu1 style] (left decoupled boundary end) to[out=100, in=-90] (left global boundary start);
  \fill[boundary insertion point, mu1color] (left decoupled boundary end) circle;
  \fill[boundary insertion point, red] (left global boundary start) circle;
\end{tikzpicture}
    \subcaption{Rescue region.}
    \label{fig:rescue-region-zoom}
  \end{subfigure}
  \begin{subfigure}{0.47\linewidth}
    \centering
    \begin{tikzpicture}[
  coupling line/.style={thick},
  mu0 line/.style={coupling line, black},
  mu1 line/.style={coupling line, mu1color},
  mu2 line/.style={coupling line, red},
  scale=1.5
  ]
  \draw[mu0 line] (0, 0) coordinate (line start) -- ++(1, 0) coordinate (mu0 end)
  node[midway, above] {$\mu = 0$};
  \path (mu0 end) -- ++(0, 1) coordinate (mu1 start);
  \path (mu1 start) -- ++(0.5, 0) coordinate (mu1 end)
  node[midway, above, mu1color] {$\mu_1$};
  \path (mu1 end) -- ++(0, -0.7) coordinate (mu2 start);
  \path (mu2 start) -- ++(1.7, 0)
  node[midway, above, red] {$\mu_2$} coordinate (mu2 end);
  \path let \p1=(mu2 end) in (\x1, 0) coordinate (line end);

  \draw[dotted] (mu0 end) -- (mu1 start);
  \draw[dotted] (mu1 end) -- (mu2 start);
  \draw[mu1 line] (mu1 start) -- (mu1 end);
  \draw[mu2 line] (mu2 start) -- (mu2 end);

  \coordinate[below=4] (axis start) at (line start);
  \coordinate[below=4] (axis end) at (line end);
  \draw[<->] (axis start) -- (axis end);

  \node[below] at (axis end) {$u$};

  \path let \p1=(mu0 end), \p2=(axis start) in
  [draw] (\x1, \y2) -- ++(0, -0.1) coordinate (minus tick end);

  \path let \p1=(mu1 end), \p2=(axis start) in
  [draw] (\x1, \y2) -- ++(0, -0.1) coordinate (plus tick end);

  \node[below] at (minus tick end) {$u_{r-}$};
  \node[below] at (plus tick end) {$u_{r+}$};
\end{tikzpicture}
    \subcaption{Simple rescue coupling.}
    \label{fig:rescue-coupling-schematic}
  \end{subfigure}
    \caption{(a) Illustration of rescued black hole geometry. The coupling $\mu(t)$ is turned on at boundary time $u_{r-}$ and by time $u_{r+}$ the boundary reaches a fixed point solution. (b) A simplest choice of $\mu(t)$ to achieve the rescued black hole geometry}
    \label{fig:rescued BH}
\end{figure}

A solution that is of particular interest to us is the ``rescued black hole'' geometry shown in Figure~\ref{fig:rescue-region-zoom}. The two SYK models remain decoupled until time $u_{r-}$. A coupling is turned on at $u_{r-}$, which provides an attractive force that pulls the boundary inwards. By fine-tuning the coupling, we can make a cross over to a fixed point solution at time $u_{r+}$. (This is just like catching an accelerating ball in an external field and holding it static.) Before time $u_{r-}$, the geometry is identical to a two-sided eternal black hole geometry, with thermal correlation functions on each boundary. Turning on the time-dependent coupling then stops the thermalization process and instead evolves the system into the ground state of a coupled model with an appropriate coupling. Consequently, the black hole horizon is removed and the regions of the bulk that would have been the black hole interior without the coupling are now accessible from the boundary. Studying this family of geometries provides a way to probe the black hole interior physics, or at least how the interior physics is modified by such rescue procedure. In the following, we will first determine the time-dependent coupling that achieves the desired solution, and then study correlation functions in this geometry.

There is a lot of freedom in choosing $\mu(t)$, which induces different rescued black hole geometries with the same initial and final state but different interpolation details. We consider a particularly simple choice (Figure~\ref{fig:rescue-coupling-schematic}) with $\mu(t)$ a step function:
\begin{align}
    \mu(t)=\left\{\begin{array}{cc}0,&u<u_{r-}\\\mu_1,&u_{r-}\leq u<u_{r+}\\\mu_2,&u\geq u_{r+}\end{array}\right.
\end{align}
If we fix $\mu_1$, the boundary will reach $p=0$ at a certain time (when the velocity in global coordinate decreases to zero). This determines $u_{r+}$. The location of the boundary $u_{r+}$ determines the value $\mu_2$, which has to be chosen such that the potential minimum is exactly at this location. Therefore the free parameters to choose are $\mu_1$ and $u_{r-}$, which determines $u_{r+}$ and $\mu_2$. Alternatively, we can also fix $u_{r-}$ and $u_{r+}$, and use them to determine the couplings $\mu_1,\mu_2$. 

The exact expressions for $u_{r+},\mu_2$ as functions of $u_{r-}$ and $\mu_1$ are derived in Appendix \ref{app:rescue}. For simplicity here we will only present results in the limit 
\begin{align}
    u_{r-}\gtrsim \csc{\epsilon} > 2,~\mu_1 > 2 \sin^2\epsilon, \label{eq:late rescue limit}
\end{align}
which means the rescue happens later than time $\sim \beta$, and the coupling is not too small (such that the rescue procedure is not too slow):
\begin{align}
    \mu_1(u_{r+} - u_{r-}) &\simeq \tan{\epsilon} \tanh(\sin \epsilon u_{r-})\label{eq:late rescue large q 1}\\
    \mu_2
    &\simeq 4\sin^2\epsilon \exp\left[-\frac{4\sin^2\frac{\epsilon}2}{\mu_1}-2\sin\epsilon u_{r-}\right].\label{eq:late rescue large q 2}
\end{align}

We are interested comparing the finite energy effective theory with the low energy Schwarzian effective theory, since the latter corresponds to the dual picture of JT gravity coupled with non-interacting matter (more precisely, matter with interaction suppressed by $\frac 1N$).  In Schwarzian theory  \eqref{eq:linear-dilaton-free-matter-action}, we can choose a rescue process that connects the thermal field double solution to the global AdS$_2$ solution within a time interval $\left[u_{r-},u_{r+}\right]$, which is qualitatively similar to the large-$q$ finite energy theory discussed above, but quantitatively different. In the same limit of Eq. (\ref{eq:late rescue limit}) the low energy theory leads to
\begin{align}
    \mu_1\left(u_{r+}-u_{r-}\right)& \simeq \sin\hat{\epsilon}\tanh\left(\sin\hat{\epsilon}u_{r-}\right)\label{eq:late rescue low energy 1}\\
    \mu_2 &\simeq 4 \sin^2 \hat{\epsilon} \exp\left[ - \frac{\sin^2 \hat{\epsilon}}{\mu_1} - 2 \sin \hat{\epsilon} u_{r-} \right]\label{eq:late rescue low energy 2}
\end{align}
with $\sin\hat{\epsilon}=\pi/\beta \mathcal{J}$.

In the limit $\beta\mathcal{J}\gg 1$, according to the definition in Eq. (\ref{eq:def epsilon}) we have $\sin\epsilon \simeq \tan\epsilon \simeq \frac{\pi}{\beta\mathcal{J}}-\frac{2\pi}{\left(\beta\mathcal{J}\right)^2}$ up to second order of $\left(\beta\mathcal{J}\right)^{-1}$. If we fix $u_{r-}$ and $u_{r+}$, and compare the $\mu_1,\mu_2$ determined by the large-$q$ theory in Eq. (\ref{eq:late rescue large q 1})-(\ref{eq:late rescue large q 2}) and that in the low energy theory in Eq. (\ref{eq:late rescue low energy 1})-(\ref{eq:late rescue low energy 2}), the deviation will be $\Delta\mu_1\propto \left(\beta\mathcal{J}\right)^{-2}$, $\Delta\mu_2\propto e^{-2 \sin\hat{\epsilon} u_{r-}} \left(\beta\mathcal{J}\right)^{-3}$. This estimation will be useful for our discussion of four-point function in the next subsection. 

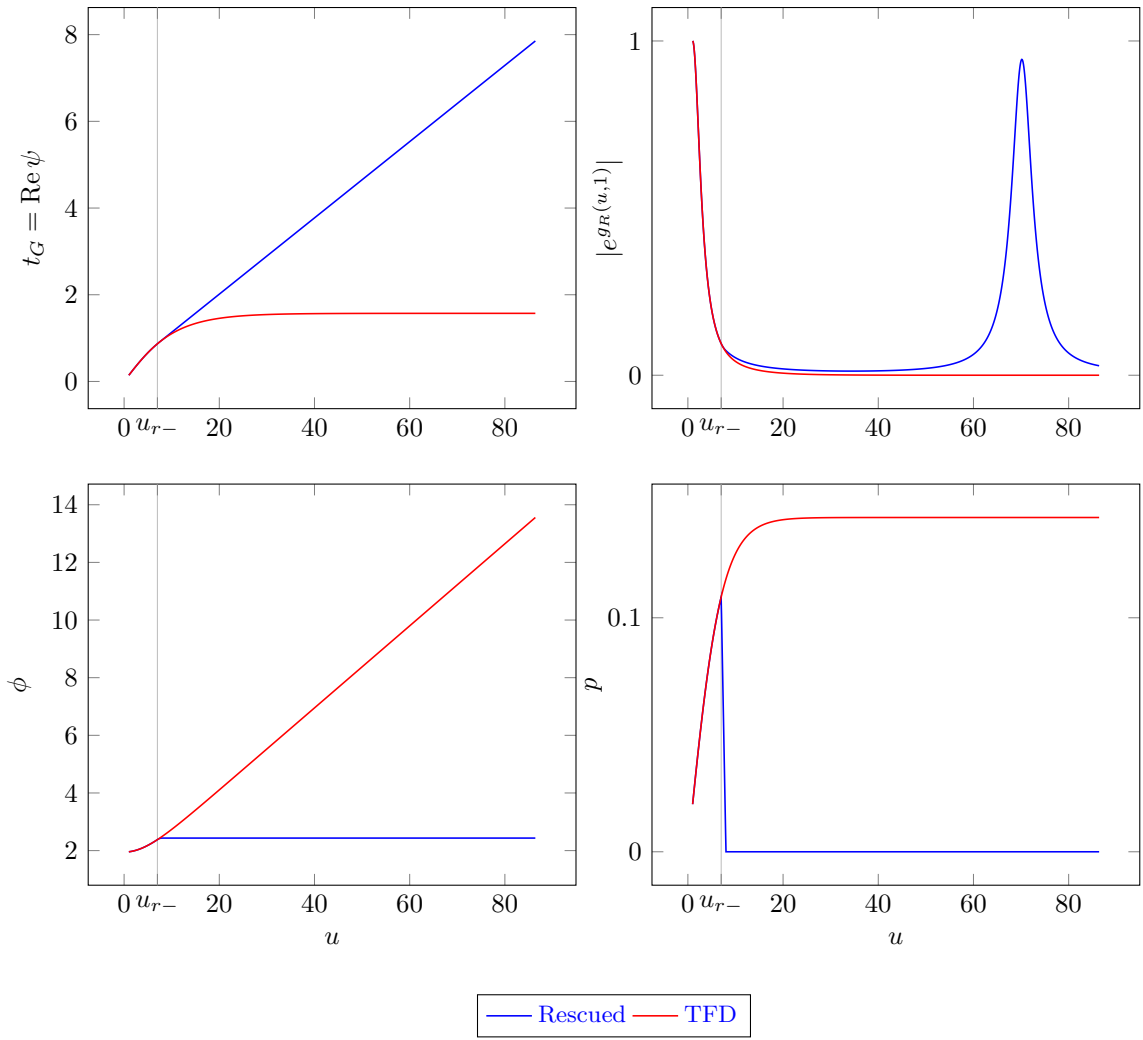
\begin{figure}
    \centering
    \begin{tikzpicture}
  \newcommand{\explotsdat}{\plotdatadir/rescued-example-psis.dat}
  \pgfplotsset{
    every axis/.append style={
      no markers},
    every axis plot/.append style={
      semithick
    }}
  \begin{groupplot}[
    group style={group size=2 by 2,
      xlabels at=edge bottom},
    xlabel={$u$},
    legend columns=2,
    extra x ticks={7},
    extra x tick style={grid=major},
    extra x tick labels={$u_{r-}$},
    width=8cm]
    \nextgroupplot[ylabel={$t_G = \Re \psi$}, legend to name=sollegend]
    \addplot table[x=u, y=resc_tG] {\explotsdat};
    \addlegendentry{Rescued}
    \addplot table[x=u, y=tfd_tG] {\explotsdat};
    \addlegendentry{TFD}
    \nextgroupplot[ylabel={$|e^{g_R(u, 1)}|$},
    ytick={0, 1},
    ylabel shift={-0.5em}]
    \addplot table[x=u, y=resc_abseg] {\explotsdat};
    \addplot table[x=u, y=tfd_abseg] {\explotsdat};
    \nextgroupplot[ylabel={$\phi$}]
    \addplot table[x=u, y=resc_phi] {\explotsdat};
    \addplot table[x=u, y=tfd_phi] {\explotsdat};
    \nextgroupplot[ylabel={$p$}, ytick={0,0.1},
    ylabel shift={-0.5em}]
    \addplot table[x=u, y=resc_p] {\explotsdat};
    \addplot table[x=u, y=tfd_p] {\explotsdat};
  \end{groupplot}
  \node at
  ($(group c1r2.south east)!.5!(group c2r2.south west) - (0, 4.5em)$)
  {\ref{sollegend}};
\end{tikzpicture}
    \caption{Plots of some simple quantities in the rescued (blue) and TFD (red) geometries. The initial TFD has $\epsilon = 1/7$ (the relationship to $\beta \mathcal{J}$ is defined in \eqref{eq:def epsilon}), and the rescue is specified by its start time, $u_{r-}=7$, and duration, $u_{r+} - u_{r-} = 1$. Starting from the upper left panel, we first show the mapping from boundary to global time $t_G$. The next panel is the absolute value of the ``self-energy'' $e^{g_R(u, u_1)}$ with $u_1=1$. Next, we show $\phi$, which as discussed in the main text is approximately (half) the geodesic distance between the boundaries. It also gives the imaginary part of the reparametrization $\Im \psi = - \tanh^{-1} e^{- \phi}$. Finally we show $p$, which for small enough $\epsilon$ has the interpretation of boundary velocity.}
    \label{fig:rescue 2pt func}
\end{figure}

Since the deviation between the low energy theory and large-$q$ theory is small in the low energy limit $\beta\mathcal{J}\gg 1$, the two point function predicted by these two theories (assuming we use the couplings $\mu_1$, $\mu_2$ as computed in that theory) will also only have a small deviation with each other. To get a sense of their behaviour, in Fig. \ref{fig:rescue 2pt func} we show the  complex reparametrization $\psi(u)$ and the rescaled self-energy $e^{g_{R}(u_2,u_1)}$ for $u_1<u_{r-}$ in the large-$q$ theory. Due to the rescue process, the exponentially decay of the self-energy in the black hole ends during the rescue and is turned into an oscillation with frequency $\omega_M$.

According to Eq. (\ref{eq:late rescue low energy 1}) and (\ref{eq:late rescue large q 2}), for $\beta\mathcal{J}\gg 1$ it is possible to rescue the black hole with small couplings $\mu_1\ll 1,~\mu_2\ll 1$. Therefore naively one would expect that the low energy Schwarzian theory is sufficient for describing the SYK physics, which would then suggest that the bulk dual theory, {\it i.e.} JT gravity coupling with 2d fermions with $1/N$ suppressed interaction, is applicable to the rescued black hole system for all time. As we will see in the next section, this is actually not true if the black hole is rescued at a late time, which can be proved at the level of four-point functions. 





\subsection{Four-point functions in the rescued geometry}
\label{sec:scatt-resc-geom}

In this section, we discuss four-point functions in the rescued geometry. In the low energy theory, we can think of these four-point functions as a probe of the back-reaction introduced by sending in a pair of shockwaves at some time $u_0$. The large-$q$ theory allows us to examine related phenomena beyond the low energy limit. 

The large-$q$ four-point functions we study are those discussed in Sec. \ref{sec:four-point-functions-1}. The four-point functions are determined by the response of the complex reparametrization field $\psi(u)$ to a perturbation at a time $u_0$. For the rescued black hole problem, the response can be computed by a Green's function method. The details of the calculation are discussed in Appendix  \ref{app:rescue 4pt}. In the main text we will only discuss the result in two special regions of time variables.

\begin{figure}[htb!]
  \centering
  \begin{subfigure}{0.63\textwidth}
    \centering
    \begin{tikzpicture}[scale=1]
  \def\boundrho{2.4}
  \def\toprindlerextent{0.1}
  \def\bottomrindlerextent{0.001}
  \def\uzinsertiongap{0.3}
  \pgfmathsetmacro\coshboundrho{0.5*(e^\boundrho + e^(-\boundrho))}
  \tikzset{declare function={
    rightrindlertG(\tG) = acos(cos(\tG r) / \coshboundrho) * pi / 180;
    leftrindlertG(\tG) = -rightrindlertG(\tG);
  }}
  \draw[conformal boundary]
  (-pi/2, -pi/2) -- (-pi/2, 3*pi/2);
  \draw[conformal boundary]
  (pi/2, -pi/2) -- (pi/2, 3*pi/2);
  \path[pattern=crosshatch, pattern color=gray!30] (0, 0) -- (pi/2, pi/2) -- (0, pi) -- (-pi/2, pi/2) -- cycle;

  \draw[light ray marker] (-pi/2, -pi/2) -- (pi/2,pi/2);
  \draw[light ray marker] (pi/2, -pi/2) -- (-pi/2,pi/2);
  \draw[light ray marker] (-pi/2,pi/2) -- ++(pi,pi);
  \draw[light ray marker] (pi/2,pi/2) -- ++(-pi,pi);

  \draw[syk boundary, name path=right rindler boundary] plot[smooth, domain=(-pi/2+\bottomrindlerextent):(pi/2-\toprindlerextent)]
  ({rightrindlertG(\x)}, \x) coordinate (right rindler end);
  \draw[syk boundary, name path=left rindler boundary] plot[smooth, domain=(-pi/2+\bottomrindlerextent):(pi/2-\toprindlerextent)]
  ({leftrindlertG(\x)}, \x) coordinate (left rindler end);

  \foreach \rindend in {left rindler end, right rindler end}
  {
    \path let \p1 = (\rindend) in [draw, syk boundary, color=red] (\p1) -- (\x1, 3*pi/2);
  }

  \draw (pi/2, {(1/sqrt(2) + 1) * pi/2}) -- ++(0.2, 0) node[right=10, rotate=90, anchor=center] {$t_G(u) \approx t_G(u_0) + \frac{\pi}{2 \sqrt{2}}$};

  \coordinate (left uz insertion) at ({leftrindlertG(pi/2-\uzinsertiongap)}, pi/2-\uzinsertiongap);
  \coordinate (right uz insertion) at ({rightrindlertG(pi/2-\uzinsertiongap)}, pi/2-\uzinsertiongap);

  \begin{scope}
    \clip let \p1=(right rindler end) in (-\x1,-pi/2) rectangle (\x1,3*pi/2);
    \draw[blue!70] (left uz insertion) -- ++(pi,pi);
    \draw[blue!70] (right uz insertion) -- ++(-pi,pi);
  \end{scope}

  \draw[boundary insertion point, fill=white] (left uz insertion) circle;
  \draw[boundary insertion point, fill=white] (right uz insertion) circle;

  \draw[boundary insertion point, fill=black]
  let \p1 = (right rindler end) in
  (\x1, {(1/sqrt(2) + 1) * pi/2}) circle;
  \draw[boundary insertion point, fill=black]
  let \p1 = (left rindler end) in
  (\x1, {(1/sqrt(2) + 1) * pi/2}) circle;
  \end{tikzpicture}
    \caption{$\mathcal{F}_{RL}(u, u, u_0)$}
    \label{fig:phi-response-4pt}
  \end{subfigure}
  \begin{subfigure}{0.32\textwidth}
    \centering
    \begin{tikzpicture}[scale=1]
  \def\boundrho{2.4}
  \def\toprindlerextent{0.1}
  \def\bottomrindlerextent{0.001}
  \def\uzinsertiongap{0.3}
  \def\u1insertiongap{0.5}
  \pgfmathsetmacro\coshboundrho{0.5*(e^\boundrho + e^(-\boundrho))}
  \tikzset{declare function={
    rightrindlertG(\tG) = acos(cos(\tG r) / \coshboundrho) * pi / 180;
    leftrindlertG(\tG) = -rightrindlertG(\tG);
  }}
  \draw[conformal boundary]
  (-pi/2, -pi/2) -- (-pi/2, 3*pi/2);
  \draw[conformal boundary]
  (pi/2, -pi/2) -- (pi/2, 3*pi/2);
  \path[pattern=crosshatch, pattern color=gray!30] (0, 0) -- (pi/2, pi/2) -- (0, pi) -- (-pi/2, pi/2) -- cycle;

  \draw[light ray marker] (-pi/2, -pi/2) -- (pi/2,pi/2);
  \draw[light ray marker] (pi/2, -pi/2) -- (-pi/2,pi/2);
  \draw[light ray marker] (-pi/2,pi/2) -- ++(pi,pi);
  \draw[light ray marker] (pi/2,pi/2) -- ++(-pi,pi);

  \draw[syk boundary, name path=right rindler boundary] plot[smooth, domain=(-pi/2+\bottomrindlerextent):(pi/2-\toprindlerextent)]
  ({rightrindlertG(\x)}, \x) coordinate (right rindler end);
  \draw[syk boundary, name path=left rindler boundary] plot[smooth, domain=(-pi/2+\bottomrindlerextent):(pi/2-\toprindlerextent)]
  ({leftrindlertG(\x)}, \x) coordinate (left rindler end);

  \foreach \rindend in {left rindler end, right rindler end}
  {
    \path let \p1 = (\rindend) in [name path=right global boundary] (\p1) -- (\x1, 3*pi/2);
  }

  \coordinate (left uz insertion) at ({leftrindlertG(pi/2-\uzinsertiongap)}, pi/2-\uzinsertiongap);
  \coordinate (right uz insertion) at ({rightrindlertG(pi/2-\uzinsertiongap)}, pi/2-\uzinsertiongap);

  \begin{scope}
    \clip let \p1=(right rindler end) in (-\x1,-pi/2) rectangle (\x1,3*pi/2);
    \draw[blue!70] (left uz insertion) -- ++(pi,pi);
    \draw[blue!70] (right uz insertion) -- ++(-pi,pi);
  \end{scope}

  \draw[boundary insertion point, fill=white] (left uz insertion) circle;
  \draw[boundary insertion point, fill=white] (right uz insertion) circle;

  \coordinate (left u1 insertion) at ({leftrindlertG(pi/2-\u1insertiongap)}, pi/2-\u1insertiongap);
  \path[name path=bg u1 prop] (left u1 insertion) -- ++(pi, pi);
  \path[name intersections={of=bg u1 prop and right global boundary}]
  (intersection-1) coordinate (right u2 insertion);

  \fill[gray!40] (left u1 insertion) -- ($(right u2 insertion) - (0, 0.1)$) -- ++(0, 0.2) -- cycle;

  \foreach \rindend in {left rindler end, right rindler end}
  {
    \path let \p1 = (\rindend) in [draw, syk boundary, color=red] (\p1) -- (\x1, 3*pi/2);
  }

  \draw[boundary insertion point, fill=black] (left u1 insertion) circle;
  \draw[fermion propagator] (left u1 insertion) -- (right u2 insertion);
\end{tikzpicture}
    \caption{Equation~\eqref{eq:4pt-interaction-diagnostic}}
    \label{fig:matter-response-4pt}
  \end{subfigure}
  \caption{Depiction of bulk interpretations of four-point functions $\mathcal{F}_{RL}(u_2, u_1, u_0)$ in the rescued black hole geometry. Open circles indicate the operator insertions at $u_0$, and dark circles on the left (right) correspond to the insertions at $u_1$ ($u_2$). We show the future light cones for $u_0$ insertions in blue lines. 
  The two subfigures correspond to the two time parameter regions we discuss. (a) corresponds to the case $u_1=u_2=u$. In (b), we schematically show the endpoints we integrate over in Eq. \eqref{eq:4pt-interaction-diagnostic} by a gray wedge.}
  \label{fig:rescued-4pt-functions}
\end{figure}
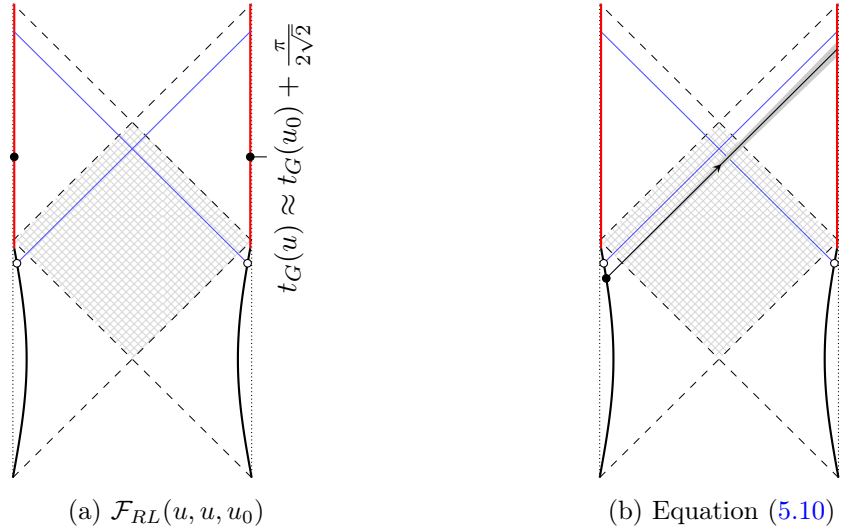

First, we consider the four-point function $\mathcal{F}_{RL}(u, u, u_0)$, which according to Eq. (\ref{eq:FRLuuu0}) is a probe of the response $\{\phi(u_0), \phi(u)\}=d\phi(u)/dp(u_0)$. 
A bulk picture for this four-point function is shown in Figure~\ref{fig:phi-response-4pt}. To give an explicit example that illustrates some general features, we consider the case $0< u_0 < u_{r-} < u_{r+} < u$, and take $u_{r-}\gg \csc\epsilon$ while holding $u_{r+} - u_{r-} \lesssim O(1)$ fixed. In this limit, the four-point function is approximately 
\begin{align}
     \mathcal{F}_{RL}(u, u, u_0)
  \sim \frac{2\tanh\left(\sin\epsilon u_0\right)}{\sin\epsilon}\sinh\left(\sin\epsilon u_{r+}\right)\sin\left(\omega_G\left(u-u_{r+}\right)\right)+\ldots\label{eq:FRLuuu0 rescue}
\end{align}
Here we have only kept the leading term that is an oscillation in $u$ with an amplitude that grows exponentially in $u_{r+}$ (a more complete expression is given in \eqref{eq:approximate-sensitivity-4pt}). The terms in $\ldots$ do not grow exponentially in this limit. Physically, we can view the response as a two-step process: the perturbation at time $u_0$ in $p(u_0)$ leads to an infinitesimal displacement in $(\phi(u_{r+}),p(u_{r+}))$, which in turn induces an oscillatory response to later time $u$ in the same way as in the global solution (\ref{eq:FRLuuu0 global}). 


The four-point function is a measure of sensitivity of the two-point function to perturbations in the coupling. The exponentially growing amplitude in Eq. (\ref{eq:FRLuuu0 rescue}) indicates that the success of the rescue ({\it i.e.} successfully landing in the global solution) becomes exponentially sensitive to the choice of coupling as rescue time $u_{r-}$ becomes large. In particular, if we use the coupling $\mu_1$ predicted by the low energy theory (\ref{eq:late rescue low energy 1}) instead of the large-$q$ theory, it is off by $\Delta\mu_1\sim (\beta\mathcal{J})^{-2}$, which will lead to a big deviation from the global geometry for large $u_{r-}$. In other words, for this four-point function, at late rescue time corrections to the low energy theory have a significant effect, but the effect can be absorbed by a correction of the coupling used. 


Now that we have shown that there are important corrections to an effective theory of JT gravity with free (to $O(1/N)$) matter, it is natural to ask what the qualitative features of a more complete gravitational description would be. One possibility is that the only change is a different dynamics of the boundary; indeed, it is exactly the response of the boundary distance that is measured by $\mathcal{F}_{RL}$. One (possibly too simplistic) approach is to test an effective description where the boundary dynamics, given by the reparametrization $t$ of the low energy theory, is replaced by the large-$q$ function $t_G = \Re \psi$, but two aspects of the JT theory are kept:
\begin{enumerate}
    \item The spacetime in the bulk (including the hatched region in Figure~\ref{fig:rescued-4pt-functions}) is locally $\ads_2$;
    \item The matter fields are minimally coupled to the metric and to leading order in $N$ do not interact with themselves or the dilaton.
\end{enumerate}
With these assumptions, the two-point function of fermions is
\begin{equation}
  \label{eq:naive-effective-two-point-function}
  G_L(u_2,u_1) \propto \left( \frac{t_G'(u_2) t_G'(u_1)}{\cos^2 \frac{t_G(u_2) - t_G(u_1)}{2}} \right)^{\Delta}.
\end{equation}
To leading order in the low energy limit, this description will give the correct two-point functions and response $\mathcal{F}_{RL}(u, u, u_0)$, even at late rescue time.

The motivation for the next four-point function we consider is to probe the consistency of this description. Intuitively, we would like to test the assumptions (1) and (2) by sending a fermion from the left to the right, and using the perturbation to create a shockwave that intersects with the fermion trajectory. Formally, the correlation function we consider is a smeared four-point function, illustrated in Figure~\ref{fig:matter-response-4pt}. We take $u_1$ and $u_2$ on opposite sides, and the time variables are ordered such that $u_1<u_0<u_2$. The near light-like separation is determined by the equation 
\begin{align}
    t_G(u_{2c}) = t_G(u_1) + \pi.
\end{align}
We are interested in how the perturbation changes the signal propagation from one boundary to the other apart from UV details, so we consider the integral of the four-point function $\mathcal{F}_{RL}(u_2, u_1, u_0)$ over an interval $u_2\in\left[u_{2c}-\frac L2,u_{2c}+\frac L2\right]$ (see Figure~\ref{fig:matter-response-4pt}).\footnote{
To be more precise, we require $t_G(u_2) - V_GL/2 > \max \{t_G(u_{r+}), t_G(u_0)\}$ such that the entire interval $u_2\in \left[u_{2c}-\frac L2,u_{2c}+\frac L2\right]$ is in the fix point region $\mu=\mu_2$, and satisfies $u_2>u_0$.} The interval size $L\ll e^{\phi(u_{r+})}$ is taken to be a finite order $1$ number, which corresponds to a time scale of order $\mathcal{J}^{-1}$ when we restore the full units. We obtain the following result:
\begin{align}
  \frac{1}{L} \Re \int_{-L/2}^{L/2} &\mathcal{F}_{RL}(u_{2c} + u, u_1, u_0) \mathrm{d}u
  \approx \left(1 - \alpha(u_1)\right) \Re A(u_{2c}, u_0) \label{eq:4pt-interaction-diagnostic}\\
  \alpha(u_1) &= \frac{4}{L} \tan^{-1} \frac{L}{2(1 + e^{ \phi(u_{r+}) - \phi(u_1)})}.
           \label{eq:alpha}
\end{align}
where we took the leading part in the low energy, late rescue time $u_{r-} \gtrsim \csc \epsilon$ limit, and also take $u_1\gtrsim \csc \epsilon$.
The coefficient $\Re A(u_{2c}, u_0)=\Re d\ln\psi'(u_{2c})/dp(u_0)$ is defined in \eqref{eq:defAB}.  

We can compare this to the low energy effective description in terms of the time $t_G$. The four point function is now determined by taking a derivative of the two-point function \eqref{eq:naive-effective-two-point-function} over a perturbation\footnote{More precisely, to carry this integration we introduced an infinitesimal cutoff by adding an $i\delta$ to $t_G(u_2)-t_G(u_1)$ and then take $\delta\rightarrow 0$.}; we find
\begin{align}
\frac{1}{L} \Re \int_{-L/2}^{L/2} &\mathcal{F}_{RL}(u_{2c} + u, u_1, u_0) \mathrm{d}u
  \approx \Re A(u_2, u_0) \label{eq:4pt-integrated-low-energy}.
\end{align}

Comparing Eq. (\ref{eq:4pt-interaction-diagnostic}) and (\ref{eq:4pt-integrated-low-energy}) we see that the actual large-$q$ integrated four-point function is smaller than the prediction of the effective low-energy description by an order one fraction $1-\alpha(u_1)$. If we fix $u_1<u_{r-}$ and increase $u_{r-}$ while keeping $u_{r+}-u_{r-}$ fixed, $\alpha(u_1)$ decays exponentially $\propto e^{-\phi(u_{r+})}$, so that the deviation between the two theories is small. However, when the initial time $u_1$ approaches $u_{r-}$, $\alpha(u_1)$ approaches a finite value $\frac{4}L\tan^{-1}\frac{L}{4}$. Technically, this discrepancy originates from the fact that the imaginary part $\Im\psi(u)$ also has a nontrivial response to the perturbation.

Physically, this result reflects the following phenomenon: for a fermion from the left boundary at a time close to the rescue time, the effect of a perturbation at later time $u_0$ to its propagation to the right-side boundary cannot be correctly predicted by an effective description assuming (1) and (2), even if the dynamics of the reparametrization $t$ is modified to be that of $\Re\psi(u)$ in the large-$q$ theory. In term of the bulk interpretation, this is suggests that 
at least one of the assumptions (1) or (2) is false. If we assume (1) is still true, then the explanation of this deviation is that the bulk fermions have a strong scattering with each other, or with the dilaton, which only becomes significant in the spacetime region near the inner horizon. We view this as a ``precursor'' for the formation of certain kind of singularity (or at least some significant deviation from the low energy gravity theory that is sensitive to the UV physics), which would have occurred if we did not rescue the black hole. 

\subsection{Finite $q$ corrections}

We close this section with some comments on finite $q$ effects. The derivation of the large-$q$ effective theory requires that 
\begin{equation}
    g_{L,R}(u_1,u_2)\ll q
\end{equation} 
for all $u_1,u_2$. Without going into details, we mention that this requirement is equivalent to requiring
\begin{equation}
    \phi(u)\ll q
\end{equation}
for all time. In the rescued black hole geometry, this requires $\phi(u_{r+})\ll q$, or equivalently
\begin{equation}
    \mu_2\gg e^{-2q}
\end{equation}
This requires that the rescue time cannot be too late:
\begin{align}
    u_{r-}\ll \csc\epsilon\left(q+\log\left(2\sin\epsilon\right)\right)\label{eq:rescue time condition 1}
\end{align}
If the rescue is later than this time, we have to return to a finite-$q$ theory and it is unclear whether it is still possible to rescue the black hole by tuning the time-dependent coupling. (On a related note, Ref. \cite{maldacena2018eternal} discussed that the ground state of the coupled SYK Hamiltonian is not exactly the thermofield double state beyond the large-$q$ limit, which suggests that even an early rescue may not be able to guide this system to the global solution.) 

Another condition that limits the applicability of our result is the size of the four-point function. The large-$N$ theory is applicable only if the four-point function satisfies
\begin{align}
    \mathcal{F}_{RS}\left(u_2,u_1,u_0\right)\ll N
\end{align}
Instead of exploring this condition systematically, we will study $\mathcal{F}_{RL}(u,u,u_0)$ in Eq. (\ref{eq:FRLuuu0 rescue}) as an example. For $u_0\gg\csc\epsilon$, the requirement is
\begin{align}
    u_{r+}\ll \csc \epsilon \log\left(N\sin\epsilon\right)\label{eq:rescue time condition 2}
\end{align}
The right-hand side of this equation is simply the scrambling time of the black hole. It should be noted that the large-$N$ large-$q$ expansion requires $N\gg q^2$, but $\log N$ could be bigger or smaller than $q$, so the two upper limits of rescue time \eqref{eq:rescue time condition 1} and \eqref{eq:rescue time condition 2} may be in either order. 


\section{Discussion and Conclusion}
\label{sec:discussion}

In summary, we have developed a general effective theory description of the coupled SYK model in the large-$N$, large-$q$ ($q^2/N\rightarrow 0$) limit. In general, we studied a quantum quench problem with the thermofield double state as the initial state, and a generic time-dependent coupling. The dynamics of this problem is mapped to a classical dynamical system. This framework allows us to study the fermion two-point functions and certain higher functions, and to gain a systematic understanding of the coupled SYK model beyond the familiar low energy limit. Compared with the low energy theory which describes a 1d reparametrization of the boundary with a Schwarzian action, the finite energy theory describes a complex reparametrization field with a modified dynamics. 

Using this general effective theory, we studied a particular situation when a two-sided black hole, corresponding to two decoupled SYK model in a time-evolved thermofield double state, is ``cooled down'' by turning on a fine-tuned time-dependent coupling. Within a finite time after the coupling is turned on, the black hole state is mapped to an eternal wormhole state, which is a thermofield double state with a lower temperature. Such a ``rescued black hole'' geometry provides a family of systems that approaches a black hole system while still keeping the entire interior available. Our effective theory allows us to detect how the bulk dual theory deviates from the low energy theory of JT gravity coupled with free fermion matter. We studied correlation functions in this geometry and show that, in addition to various quantitative corrections to the low energy theory, certain 4-point functions obtain leading order corrections due to the finite energy effect. Physically, this indicates that the physics near the inner horizon region deviates from the low energy theory even if the initial temperature and the coupling are always small. We consider this as an indication of the singularity that would have been formed if we had not rescued the black hole.


There are many open questions remaining in this setting. A very natural one is to construct a bulk dual theory for the large-$q$ finite energy theory. Corrections to the JT gravity theory have been discussed before ({\it c.f.} \cite{kitaev2019statistical,witten2020deformations}) but no dual theory has been derived for the SYK model beyond the low energy limit (to all orders in $\beta \mathcal{J}$ and $\mu$). The Liouville effective theory in large-$q$ limit suggests that a dual theory that is simpler than the finite $q$ SYK model might be possible. Mapping the Liouville theory to a bulk dual theory (which correctly reduces to the JT gravity in the appropriate low energy limit) is the topic of our ongoing work\cite{lensky2021holographic}.

Another question is the relation between the rescued black hole geometry and the entanglement island phenomena\cite{penington2020entanglement,almheiri2019entropy}. If we couple the AdS black hole with a bath, the entanglement entropy between the bath and the black hole experiences a Page phase transition, after which time the entanglement wedge of the bath includes a region in the interior of black hole, known as the (causal wedge of the) entanglement island. Ref. \cite{almheiri2019islands} studied the case of an eternal two-sided black hole coupled with thermal bath, and Ref. \cite{penington2019replica,chen2020replica} studied this phase transition in an SYK model coupled with thermal baths. Since the entanglement island also provides a way to probe the black hole interior, it is interesting to ask whether there is any relation between the island (and the replica wormhole in Renyi entropy calculation) and the rescued black hole geometry.

\noindent{\bf Acknowledgement.} We would like to acknowledge helpful discussions with Felipe Hernandez, Juan Maldacena and Ying Zhao. This work is supported by the National Science Foundation Grant No. 1720504 (YL and XLQ), the Simons Foundation (XLQ) and the Hertz Foundation (YL). This work is also supported in part by the DOE Office of Science, Office of High Energy Physics, the grant de-sc0019380.

\bibliography{refs,syk_bulk_singularity}
\bibliographystyle{jhep}

\appendix

\section{Direct derivation of the Hamiltonian}
\label{sec:direct-deriv-hamilt}

In this appendix, we derive the ODE system for the large-$q$ two-point function of the quenched thermofield double. We write $\partial_{\tau} = \partial$, $\partial_{\overline{\tau}} = \overline{\partial}$. Using the definition $G(\tau, \overline{\tau}) = G_0(\tau, \overline{\tau})(1 + g(\tau, \overline{\tau}) / q) = \frac{1}{N} \sum_{j=1}^N \chi_j(\tau) \chi_j(\overline{\tau})$, where
\begin{equation}
  \chi_j(\tau) =
  \begin{cases}
    \chi_{R j}(\tau) & \Im \tau > 0 \\
    i \chi_{L j}(- \tau) & \Im \tau < 0
  \end{cases},
\end{equation}
we have the operator relations (setting the left argument of $g$ to have larger imaginary part than the right argument)
\begin{align}
  \label{eq:anticommutator-constraint}
  \frac{1}{2} (\partial + \overline{\partial}) g(\tau, \tau)
  &= \frac{q}{2N} \sum_{j=1}^N \frac{d}{d\tau} \{\chi_j(\tau), \chi_j(\tau)\}
    = 0 \\
  \frac{1}{2} (\partial - \overline{\partial}) g(\tau, \tau)
  \label{eq:syken-constraint}
  &= \frac{q}{N} \sum_{j = 1}^N \left[ \frac{d \chi_j(\tau)}{d\tau}, \chi_j(\tau) \right]
    = \frac{2 q^2}{N} H_{\text{SYK}}^{(R/L)} - 2 \mu(\tau) G(\tau, -\tau) \\
  \label{eq:syken-derivative}
  \frac{\mu(\tau)}{2} (\partial - \overline{\partial}) g(\tau, -\tau)
  &= \frac{d}{d\tau} \left( \frac{q^2}{N} (H_{\text{SYK}}^L + H_{\text{SYK}}^{R}) \right).
\end{align}
From the large-$q$ Liouville boundary condition, we also find
\begin{equation}
  \label{eq:non-operator-gtmt-bc}
  \frac{1}{2} (\partial + \overline{\partial}) g(\tau, -\tau) = - \mu(\tau).
\end{equation}
The condition \eqref{eq:non-operator-gtmt-bc} also arises as the leading part of the operator relation
\begin{equation}
  \label{eq:operator-gtmt-bc}
  \frac{1}{2} (\partial + \overline{\partial}) g(\tau, -\tau) =
  - \mu(\tau) + i \frac{q}{N} (H_{\text{SYK}}^{\tilde{L}} + H_{\text{SYK}}^{\tilde{R}})
\end{equation}
where $H_{\text{SYK}}^{\tilde{S}}$ is obtained by taking the Hamiltonian $H_{\text{SYK}}^S$ for the appropriate side, and for each term generating $q$ terms by replacing one of the Majorana operators with the corresponding one from the opposite side. To find the real-time conditions we just take $\tau \to i t$.

We take expectation values of the above operator equations in the thermofield double state, take the leading part in $q$~\footnote{Actually, to get a similar problem, we really only need the left-right symmetry in $H_{\text{SYK}}^{L/R}$, that the correction in \eqref{eq:operator-gtmt-bc} is small, and that $\partial_{\tau} G(\tau, -\tau) \ll 1$ (then we can renormalize $\mu$ by scaling it by $2 G(\tau, -\tau)$).}, and define $\varepsilon_{\text{SYK}} = 2 q^2 \langle H_{J \chi_R}(t) \rangle / N$. We define, for $t, t', \tau \in \mathbb{R}$, $t > t'$, $|\tau| \le \beta / 4$, $g_R(t, t') = g(\beta/4 + i t, \beta / 4 + it')$, $g_L(t, t') = g(\beta / 4 + i t, - \beta / 4 - i t')$, and $g_I(t, \tau) = g(\beta / 4 + it, \tau)$. When we wish to keep the subscript free we write $g_S$, $S \in \{R, L, I\}$. The relations \cref{eq:anticommutator-constraint,eq:syken-constraint} apply to $g_R$, and \cref{eq:syken-derivative,eq:non-operator-gtmt-bc,eq:operator-gtmt-bc} apply to $g_L$. Furthermore, each of $g_L$, $g_R$, and $g_I$ solve a Liouville equation from the appropriate factors of $\pm i$ in \eqref{eq:large-q-concise-equation}.

To find the form of the general solution of these equations applicable to our case, we use arguments similar to ref. \autocite{tsutsumi1980solutions} (our case is slightly more complicated since we have to relate $g_L$, $g_R$, and $g_I$ and we are interested in the case that $\mu$, and hence our solution, is not analytic). We take $\mu$ to be twice continuously differentiable, or approximate $\mu$ by such functions and take a limit. Then all $g_S$ are 3 times differentiable on the interior of their domains. Furthermore, we assume there is some nonzero amount (which can be taken arbitrarily small) of Lorentzian time evolution before $\mu > 0$\footnote{We can actually take the more general assumption that $\mu$ is analytic in imaginary time, and there is a nonzero amount of Lorentzian time before it becomes non-analytic.}. Then $g(\beta / 4 + i t, \overline{\tau})$ and its derivatives are analytic in $\overline{\tau}$ in some strips centered on $\overline{\tau} = \pm \beta / 4$ for all $t$ even after the coupling is turned on. There is a ``conserved'' quantity\footnote{These are just proportional to the nonzero components of the stress tensor of a Liouville CFT.}
\begin{align}
  T[\hat{g}(x, y)] &= \partial_x^2 \hat{g}(x, y) - \frac{1}{2} (\partial_x \hat{g}(x, y))^2,\; \partial_y T[\hat{g}(x, y)] = 0 \\
  \overline{T}[\hat{g}(x, y)] &= \partial_y^2 \hat{g}(x, y) - \frac{1}{2} (\partial_y \hat{g}(x, y))^2,\; \partial_x \overline{T}[\hat{g}(x, y)] = 0
\end{align}
where the equations hold when $\hat{g}$ satisfies a Liouville equation with any nonzero complex coefficient $\lambda$ in front of the exponential, $\partial_x \partial_y \hat{g}(x, y) = - 2 \lambda e^{\hat{g}(x, y)}$. The Liouville equation has a ``conformal symmetry'', in that for any solution $\hat{g}(x, y)$, the function
\begin{equation}
\hat{g}_{(f,\overline{f})}(x, y) = \ln \partial_x f(x) + \ln \partial_y \overline{f}(y) + \hat{g}(f(x), \overline{f}(y))
\end{equation}
is also a solution. We also call this a ``reparametrization'' of $x$ or $y$. In this section, a subscript with functions in the parentheses means to reparameterize the left coordinate by the first function, and the right coordinate by the second in this way. If we omit the second function, it is implied to be the identity. Under this conformal transformation, we have
\begin{equation}
  \label{eq:liouville-stress-tensor-transformation}
  T[\hat{g}_{(f, \overline{f})}] = \left. T[\hat{g}]\right|_{f(x)} (\partial_x f)^2 + \{f, x\}, \;
  \overline{T}[\hat{g}_{(f, \overline{f})}] = \left. \overline{T}[\hat{g}]\right|_{\overline{f}(x)} (\partial_y \overline{f})^2 + \{\overline{f}, y\}.
\end{equation}
One way to see this is to note that if we have a function $\hat{G}(x, y)$ with $\hat{g}(x, y) = \ln \partial_x \hat{G}(x, y) + \tilde{G}(y)$ for an arbitrary $\tilde{G}$, we have $T[\hat{g}] = \{\hat{G}, x\}$, and when we reparameterize $\hat{g}$ we can just take $\hat{G}(x, y) \to \hat{G}(f(x), \overline{f}(y))$. Then we just use the composition rule of the Schwarzian $\{f \circ \hat{g}, x\} = \{f, \hat{g}\} (\partial_x \hat{g})^2 + \{\hat{g}, x\}$.

In particular, given some solution $\hat{g}(x,y)$, we can solve for an $f_P(x)$ such that $T[\hat{g}] = \{f_P, x\}$, and likewise for $\overline{f}_P(y)$. We find
\begin{equation}
  0 = \{f_P \circ f_P^{-1}, x\} = T[\hat{g}_{(f_P^{-1}, \overline{f}_P^{-1})}],
\end{equation}
(where we used \eqref{eq:liouville-stress-tensor-transformation}) and likewise $\overline{T}[\hat{g}_{(f_P^{-1}, \overline{f}_P^{-1})}] = 0$. All complex $\hat{g}$ solving $T[\hat{g}] = \overline{T}[\hat{g}] = 0$ and the Liouville equation with constant $\lambda$ are: $g_P(x, y) = - 2 \ln (x - y) - \ln \lambda + i 2 \pi n$ for $n \in \mathbb{Z}$, along with all $\text{SL}(2; \mathbb{C})$ reparametrizations of $x$ and $y$ ($\text{SL}(2; \mathbb{C})$ acts as the Mobius transformations). That this is the most general solution can be easily seen from solving $T[\hat{g}] = 0 \implies \partial_x \hat{g} = - 2 / (x - F(y))$ for some function $F$, and likewise for $\overline{T}[\hat{g}] = 0$. The $\text{SL}(2; \mathbb{C})$ reparametrizations are present because the Schwarzian derivative vanishes exactly on the Mobius transformations. Since the $f_P$ (and likewise the $\overline{f}_P$) are only determined up to an action of $\text{SL}(2; \mathbb{C})$, we find that all solutions can be written as a reparametrization of $g_P$. Results of this form are well-known (c.f.~\autocite{tsutsumi1980solutions, liouville1853equation}).

Thus the known results are enough to show that for each of the $g_S$, there is a local expression in terms of a reparametrization of $g_{P}$ (the phase is fixed by the boundary conditions and continuity) by $f_S$ and $\overline{f}_{S}$ for the left and right coordinates, respectively. The point of going through the above discussion in detail is to show that we can take $f_R = f_L = f_I$. Consider the equality $f_R = f_I$; $f_L = f_I$ is completely analogous. Then our above discussion shows that $T[g_{I,(f_I^{-1})}] = 0$, and our analyticity assumptions show that in some small $t'$ interval we can take $g_R(t, t') = g_I(t, \beta / 4 + i t')$. Thus in this strip we can take $f_I = f_R$ (possibly by precomposing with some $\text{SL}(2; \mathbb{C})$), and we can extend past the strip in $t'$ by noting $\partial_{t'} T[g_{R, (f_I^{-1})}] = 0$ in the interior of the domain of $g_R$, so $T[g_{R, (f_I^{-1})}] = 0$ on the interior of the domain of $g_R$.

Therefore, to find the real-time two-point functions we just need to solve for the four functions $f_R$, $\overline{f}_R$, $\overline{f}_L$, and $\varepsilon_{\text{SYK}}$. The Liouville equation is automatically solved, and we just need to solve the four constraints \cref{eq:anticommutator-constraint,eq:syken-constraint,eq:syken-derivative,eq:non-operator-gtmt-bc}. This is a system of four ODEs for four functions, so the solution is unique subject to initial conditions on the values and derivatives of the reparametrizations, and on the value of $\varepsilon_{\text{SYK}}$~\footnote{We are implicitly using the fact that we know the solution before $\mu > 0$, and whenever $\mu = 0$, $\varepsilon_{\text{SYK}}$ is a constant.}. It is convenient to instead solve for $\psi$, $\overline{\psi}_R$, and $\overline{\psi}_L$, defined by 
\begin{align}
    f_R(t) = \tan (\psi(t) / 2),~\overline{f}_R(t) = \tan (\overline{\psi}_R(t) / 2),~\overline{f}_L(t) = - \cot(\overline{\psi}_L(t) / 2).
\end{align} 
For convenience, we note that if we define
\begin{align}
  g_{0R}(t, t') &= - 2 \ln \sin \frac{t - t'}{2} - i \pi \\
  g_{0L}(t, t') &= - 2 \ln \cos \frac{t - t'}{2}
\end{align}
then the two-point functions are just $g_R = g_{0R,(\psi_R, \overline{\psi}_R)}$, $g_L = g_{0L, (\psi_R, \overline{\psi}_L)}$. These can then be substituted into the constraints to find the differential system for the reparametrizations and $\varepsilon_{\text{SYK}}$.

The thermofield double initial conditions are such that we have $\psi(t_0) = \overline{\psi}_R^{*}(t_0) = \overline{\psi}_L^{*}(t_0)$, with the derivatives matching as well. We could argue that since the evolution can be made analytic with arbitrarily small $\mu$ (or just try constant $\mu$), these conditions are consistent with the constraints algebraically if they are consistent at some time, and we can just take $\psi = \overline{\psi}_R^{*} = \overline{\psi}_L^{*}$. More directly, we simply substitute the ansatz $\psi = \overline{\psi}_R^{*} = \overline{\psi}_L^{*}$, find $\varepsilon_{\text{SYK}}(t)$ in terms of $\psi$ and $\psi'$ using \eqref{eq:syken-constraint}, and check that this is consistent with \eqref{eq:syken-derivative} when we implement the other 2 constraints~\footnote{The most convenient way to check is to first assume $\Re \psi' \ne 0$ (we will always assume, and see later that it is self-consistent for our cases of interest, that $|\psi'| \ne 0$ and $0 < |\Im \psi| < \infty$), then check the case $\Re \psi' = 0$ separately.}. Then we can just use \eqref{eq:anticommutator-constraint} and \eqref{eq:non-operator-gtmt-bc} to determine $\psi$, eliminating the $\varepsilon_{\text{SYK}}$ function altogether.

Thus we have reduced our problem to a system of differential equations for $\psi$. The equation \eqref{eq:anticommutator-constraint} can be integrated immediately to give
\begin{equation}
  \label{eq:magnitude-psi-derivative-generic}
  |\psi'| = c_0 \sinh |\Im \psi|
\end{equation}
for some integration constant $c_0 > 0$. Since this is a constant, it can just be evaluated on the initial conditions. Furthermore, by rescaling time we also rescale $c_0$; another way to say this is that the reparametrization $\psi$ is dimensionless, so $c_0$ has dimensions of inverse time, and can by a choice of unit be set to $1$. Our initial conditions are from the thermofield double $\mu=0$ solutions. The appropriate reparametrization $\psi$ can be easily found from the known solution~\autocite{maldacena2016remarks}, where the constants $\epsilon$ and $v$ are defined by $\sin \epsilon = \pi v / (\beta \mathcal{J})$ and $\epsilon = \pi(1 - v) / 2$. We display the reparametrization both in the more usual time $\tilde{u} = 2 \pi t / \beta$, and $u = v \csc \epsilon \tilde{u} = 2 \mathcal{J} t$:
\begin{align}
  \psi(\tilde{u}) &= 2 \tan^{-1} \left( \tanh \left( \frac{v \tilde{u} - i \epsilon}{2} \right) \right)
                    = 2 \tan^{-1} \left( \tanh \left( \frac{\sin (\epsilon) u - i \epsilon}{2} \right) \right) \\
  \psi'(u) &= \sin \epsilon \sech (\sin (\epsilon) u - i \epsilon) \\
  - \tanh \Im \psi(u) &= \sin \epsilon \sech (\sin (\epsilon) u)
  , \; |\psi'(u)| = \frac{\sin \epsilon \sech (\sin (\epsilon) u)}{\sqrt{1 - \sin^{2} \epsilon \sech^2 (\sin (\epsilon) u)}}
\end{align}
By substituting these equations into~\eqref{eq:magnitude-psi-derivative-generic}, we find if we use the time $u$, $c_0 = 1$.

We then define $K = \mu / (2 \mathcal{J})$ and call the phase of $\psi'$, $p$ ($\psi' = |\psi'|e^{i p}$). The condition \eqref{eq:non-operator-gtmt-bc} leads to the equation
\begin{equation}
  p' = - K + \frac{|\psi'|^2}{\sqrt{1 + |\psi'|^2}} \cos p = - K + \sinh (\Im \psi) \tanh (\Im \psi) \cos p.
\end{equation}
Therefore we have a natural phase space in terms of $|\psi'|$ and $p$.

It is useful to note that this ODE system arises from a Hamiltonian. In terms of the variable $\phi = \frac{1}{2} \ln (1 + |\psi'|^{-2})$, we have the Hamiltonian
\begin{equation}
  \lgqeham = - \sqrt{1 - e^{- 2 \phi}} \cos p + K \phi
\end{equation}
with canonical coordinate $\phi$ and canonical momentum $p$. In the case $K = 0$, we find solutions
\begin{align}
  \label{eq:decoupled-canonical-variables}
  \phi(u) &= - \ln \sin \epsilon + \ln \cosh (\sin (\epsilon) u) \\
  p(u) &= \tan^{-1} \left( \tanh(\sin (\epsilon) u ) \tan \epsilon \right)
\end{align}
where $\epsilon$ is an integration constant.

\section{Rescued black hole solution}
\label{app:rescue}

In this appendix, we give explicit and practical expressions for finding a sequence of couplings to achieve the rescue geometry described in Section~\ref{sec:rescue}.

Suppose just before time $u_{r-}$ the system has energy $\lgqeham{}_{r-}$, and coordinates $\phi_{r-}$ and $p_{r-}$. Then when we turn on coupling $\mu_1$, the system has a new energy $\lgqeham{}_{r-} + \mu_1 \phi_{r-}$. Calling $\delta \phi = \phi_{r+} - \phi_{r-}$, we have the fixed point equation
\begin{align}
  \delta \phi &= \frac{\sqrt{1 - e^{- 2 \phi_{r-}} e^{- 2 \delta \phi}} + \lgqeham{}_{r-}}{\mu_1} \\
  &= \frac{2 \sin^2 \frac{\epsilon}{2} - \frac{e^{- 2 \phi_{r-}} e^{- 2 \delta\phi}}{1 + \sqrt{1 - e^{- 2 \phi_{r-}} e^{- 2 \delta \phi}}}}{\mu_1}
\end{align}
where we also wrote the explicit expression for the case of Rindler initial conditions. In general, there are two solutions of this fixed point equation. We focus on the solution with $\delta \phi > 0$, which from the bulk point of view means that the boundaries are never closer to each other than they started at $u_{r-}$. The second coupling is then determined in terms of $\delta \phi$,
\begin{equation}
  \mu_2 = \frac{e^{- 2 \phi_{r+}}}{\sqrt{1 - e^{- 2 \phi_{r+}}}}.
\end{equation}
If $\delta \phi > 0$ ($\delta \phi < 0$) then $\mu_2 < \mu_1$ ($\mu_2 > \mu_1$). Finally, the rescue time in the case $\delta \phi > 0$ is
\begin{align}
  u_{r+} - u_{r-}
  &= \int_{0}^{\delta \phi} \frac{1}{\sqrt{1 - e^{- 2 \phi_{r-}} e^{- 2 \phi} - (\mu_1 \phi - \lgqeham{}_{r-})^2}} d\phi \\
  \begin{split}
    &=
    \frac{\tan \epsilon \tanh (\sin \epsilon u_{r-})}{\mu_1 - \tan \epsilon \sin \epsilon \sech^2 (\sin \epsilon u_{r-})} \\
    &\quad - \int_0^{\delta \phi} \sqrt{1 - e^{- 2 \phi_{r-}} e^{- 2 \phi} - (\mu_1 \phi - \lgqeham{}_{r-})^2}
    \frac{2 e^{- 2 \phi_{r-}} e^{- 2 \phi} + \mu_1^2}{(e^{- 2 \phi_{r-}} e^{- 2 \phi} - \mu_1(\mu_1 \phi - \lgqeham{}_{r-}))^2} d\phi.
  \end{split}
\end{align}

Let us consider the two-point function in this ``rescued'' geometry. When both coordinates are before or after the rescue, we have discussed the two-point functions above; they are just the ``regulated'' Rindler and global two-point functions. Another case is the two-point function $G_S(u_2, u_1)$ with $u_1 < u_{r-} < u_{r+} < u_2$. We just need the form of $\psi$ in the regions of interest,
\begin{equation}
  \psi(u) =
  \begin{cases}
    2 \tan^{-1} \tanh \frac{\sin (\epsilon) u - i \epsilon}{2} & u \le u_{r-} \\
    V_G (u - u_{r+}) - i \tanh^{-1} e^{- \phi_G} + V_R(u_{r+} - u_{r-}) + \Re \psi(u_{r-}) & u > u_{r+}
  \end{cases}
\end{equation}
where $V_R$ is defined by
\begin{equation}
  V_R = \frac{1}{u_{r+} - u_{r-}} \Re \int_{u_{r-}}^{u_{r+}} \psi'(u) du \approx \frac{1}{2} \sec \epsilon \csch \phi_{r-}.
\end{equation}
We can think of $V_R$ as the average rate of global time advance during the rescue.

We can repeat the above discussion in the low energy theory \eqref{eq:linear-dilaton-free-matter-action} in the large $q$ limit. Here we find
\begin{equation}
  \hat{\phi}_{r+} - \hat{\phi}_{r-} = \delta \hat{\phi}
  = \frac{\frac{\sin^2 \hat{\epsilon}}{2} - \frac{1}{2} e^{- 2 \hat{\phi}_{r-}} e^{- 2 \delta \hat{\phi}}}{\mu_1}.
\end{equation}
The second coupling is just $\mu_2 = e^{- 2 \hat{\phi}_{r+}}$. The rescue time interval is
\begin{align}
  u_{r+} - u_{r-} &= \int_0^{\delta \hat{\phi}} \frac{d \phi}{\sqrt{\sin^2 \hat{\epsilon} - e^{- 2(\hat{\phi}_{r-} + \phi)} - 2 \mu_1 \phi}} \\
  \begin{split}
                  &= \frac{\sin \hat{\epsilon} \tanh(\sin (\hat{\epsilon}) u_{r-})}{\mu_1 - \sin^2 \hat{\epsilon} \sech^2 (\sin (\hat{\epsilon}) u_{r-})} \\
                    & \quad - 2 \int_0^{\delta \hat{\phi}} \sqrt{\sin^2 \hat{\epsilon} - e^{- 2(\hat{\phi}_{r-} + \phi)} - 2 \mu_1 \phi} \frac{e^{- 2 (\hat{\phi}_{r-} + \phi)}}{(e^{- 2(\hat{\phi}_{r-} + \phi)} - \mu_1)^2}.
                  \end{split}
\end{align}

\section{Details on the four-point function computation}
\label{sec:details-four-point}

\subsection{Response to coupling perturbations}
\label{sec:response}

In this section, we study some higher-order features of the dynamics by looking at the response of the system to perturbation in the coupling $\mu$. The main application is to compute certain higher order correlation functions; we will discuss the four-point function case in detail in Section~\ref{sec:four-point-functions}. Consider then some perturbation $\mu_{\nu}(u) = \mu(u) + \nu \delta(u - u_0)$. Of course, before $u = u_0$, there is no response. Immediately after $u = u_0$, from the equations of motion for $p$ we have $p(u_0^{+}) \to p(u_0) - \nu$, while $\phi(u_0^{+}) \to \phi(u_0)$. Thus the $\delta$-function perturbation to the coupling $\mu(u)$ corresponds to an impulse that shifts the momentum. The responses $d \phi(u) / d\nu$ and $dp(u) / d\nu$ at $\nu = 0$ are therefore determined by $- \xi(u)$, with %
\begin{equation}
  \xi(u) = \begin{pmatrix} \frac{d\phi(u)}{dp(u_0)} \\ \frac{dp(u)}{dp(u_0)} \end{pmatrix}
  , \; \xi(u_0) = \begin{pmatrix} 0 \\ 1 \end{pmatrix}.
\end{equation}
The two components of $\xi$ are just the Poisson brackets $\{\phi(u_0), \phi(u)\}$ and $\{\phi(u_0), p(u)\}$.

To compute $\xi(u)$, we take derivatives of the equations of motion to get a coupled first order system,
\begin{align}
  M(\phi, p, u)
  &= \frac{\partial^2 \lgqeham}{\partial \phi \partial p} \sigma^z + \frac{\partial^2 \lgqeham}{\partial p^2} \sigma_+
      - \frac{\partial^2 \lgqeham}{\partial \phi^2} \sigma_- \\
  \frac{d \xi(u)}{du} &= M(\phi, p, u) \xi(u)
\end{align}
where $\sigma_{\pm} = \frac{1}{2}(\sigma^x \pm i \sigma^y)$. The equations can be simplified by a conjugation.
Define
\begin{align}
S(u)&= \left( \sqrt{1 - e^{- 2 \phi(u)}} \right)^{-\sigma^z} =\left(\begin{array}{cc}
   1/\sqrt{1-e^{-2\phi(u)}}  &  0\\
    0 & \sqrt{1-e^{-2\phi(u)}} 
\end{array}\right)\\
  \xi_1(u) &\equiv S(u)\xi(u) \\
  \frac{d\xi_1(u)}{du}&=M_1(\phi,p,u)\xi_1(u)\\
  M_1(\phi, p, u) &= SMS^{-1}-S\frac{dS^{-1}}{du}=\frac{\mu \phi - \lgqeham}{1 - e^{- 2 \phi}}
                 \begin{pmatrix}
                   0 & 1 \\ - e^{- 2 \phi} (2 - e^{- 2 \phi}) & 0
                 \end{pmatrix} 
\end{align}
We can define the Green's function for $\xi_1(u)$ as
\begin{align}
\xi_1(u)&=G_1(u,u')\xi_1(u')\\
  \label{eq:G1-matrix-def}
  \frac{d G_1(u, u')}{du} &= M_1(\phi, p, u) G_1(u, u'), \; G_1(u, u) = \idop.
\end{align}
Note that $M_1$ is always conjugate to a matrix proportional to $i \sigma^y$, and $G_1 \in \text{SL}(2, \mathbb{R})$. Therefore finding the response amounts to finding the matrix $G_1(u, u') \in \text{SL}(2, \mathbb{R})$. (In order to patch together known solutions using $G_1(u, u'') G_1(u'', u') = G_1(u, u')$, we will need the whole matrix.) It is convenient to define
\begin{equation}
  \tilde{\xi}(u) \equiv G_1(u, u_0) \xi(u_0)
\end{equation}
so that
\begin{equation}
  \xi(u) = \sqrt{1 - e^{- 2 \phi(u_0)}} \left( \sqrt{1 - e^{- 2 \phi(u)}} \right)^{\sigma^z} \tilde{\xi}(u).
\end{equation}
where we have used $\xi(u_0)=(0~1)^T$.


For general time dependent coupling, Eq. (\ref{eq:G1-matrix-def}) can be integrated numerically. As two simple examples we find the $G_1$ matrix analytically for the global solution and the decoupled solution. The global solution is the simplest case since $M_1$ is constant due to time-translation symmetry. In this case, $M_1$ is conjugate to a constant times $i \sigma^y$. To find the coefficient, we note that $M_1$ is conjugate to $M$ when $\phi$ is constant, which has the same determinant as the Hessian of $\lgqeham$; this last determinant is just the harmonic frequency at the fixed point $\omega_G$. Thus the solution is
\begin{align}
  \label{eq:global-G1-matrix}
  G^{(G)}_1(u) &= e^{(\phi_G/2 - \frac{1}{4} \ln (2 - e^{- 2 \phi_G})) \sigma^z} e^{i \omega_G u \sigma^y} e^{-(\phi_G/2 - \frac{1}{4} \ln (2 - e^{- 2 \phi_G})) \sigma^z} \\
         &= \cos (\omega_G u) +
           \begin{pmatrix}
             0 & \frac{e^{\phi_G}}{\sqrt{2 - e^{- 2 \phi_G}}} \\
             - e^{- \phi_G} \sqrt{2 - e^{- 2 \phi_G}} & 0
           \end{pmatrix} \sin (\omega_G u) \\
  G^{(G)}_{1}(u, u') &= G^{(G)}_{1}(u - u').
\end{align}
We point out that there is exponential growth in the response $\{\phi(u_0), \phi(u)\}$ as a function of the fixed point value $\phi_G$. 

The decoupled case $\mu = 0$ is more complicated (it can be found by solving the second order differential equation arising from the first order system)
\begin{align}
  G^{(R)}_1(u)
  &= e^{(\phi / 2) \sigma^z} e^{i t_G(u) \sigma^y} e^{(\phi / 2) \sigma^z} e^{- B_- \cos \epsilon u} e^{\ln \sin \epsilon \sigma^z} \\
  &= \frac{1}{\sqrt{1 - e^{- 2 \phi}}}
    \begin{pmatrix}
      \cot \epsilon & \tanh (\sin (\epsilon) u) \\
      - \tanh (\sin (\epsilon) u) & e^{- 2 \phi} \cot \epsilon
    \end{pmatrix} e^{- B_- \cos \epsilon u} e^{\ln \sin \epsilon \sigma^z} \\
  \tilde{\xi}(u) &= G^{(R)}_1(u, u_0) \xi(u_0)
                 = G^{(R)}_1(u) G^{(R)}_1(u_0)^{-1}\xi(u_0) \\
  \begin{split}
  &= \frac{\cot \epsilon}{\sqrt{(1 - e^{- 2 \phi(u)})(1 - e^{- 2 \phi(u_0)})}} \\
  &\qquad
    \begin{pmatrix}
      \tanh(\hat{u}) - \tanh (\hat{u}_0) + (\hat{u} - \hat{u}_0) \tanh (\hat{u}) \tanh (\hat{u}_0) \\
      \tan \epsilon \tanh(\hat{u}) \tanh(\hat{u}_0) + \cot \epsilon e^{- 2 \phi(u)} (1 + (\hat{u} - \hat{u}_0) \tanh (\hat{u}_0))
    \end{pmatrix},
  \end{split}
\end{align}
where we used a rescaled time $\hat{u} = \sin (\epsilon) u$. This gives the response of $\phi$ and $p$ in the $\mu = 0$ system,
\begin{align}
  \{\phi(u_0), \phi(u)\} &= \cot \epsilon \left( \frac{\sinh (\hat{u} - \hat{u}_0)}{\cosh (\hat{u}) \cosh (\hat{u}_0)} + (\hat{u} - \hat{u}_0) \tanh (\hat{u}) \tanh (\hat{u}_0) \right) \\
  \{\phi(u_0), p(u)\} &= \frac{1}{1 - e^{- 2 \phi(u)}} \left( \tanh(\hat{u}) \tanh(\hat{u}_0) + \cos^2 (\epsilon) \sech^{2}(\hat{u}) (1 + (\hat{u} - \hat{u}_0) \tanh (\hat{u}_0)) \right).
\end{align}

Note that if we have a solution that is in a decoupled orbit for $u \le u_{r-}$, in some unknown orbit for $u_{r-} < u \le u_{r+}$, and then in a fixed point orbit for $u_{r+} < u$, then we only have to solve for $G_1(u, u_{r-})$ for $u_{r-} < u \le u_{r+}$. Then we can find the entire $G_1(u, u_0)$ matrix, and hence the response $\xi(u)$, by composition with the known solutions $G^{(R)}_1$ and $G^{(G)}_1$. This will be discussed in the next subsection for the rescued black hole geometry.

\subsection{Four-point functions from response theory}
\label{sec:four-point-functions-response-app}

We find the four-point function by varying the two-point function with respect to the perturbation $\mu_{\nu}(u) = \mu(u) + \nu \delta(u - u_0)$ considered in Section~\ref{sec:response},
\begin{align}
  \left. \frac{d g_{S}(u_2, u_1)}{d \nu} \right|_{\nu = 0}
  &=
    \begin{split}
    - \frac{2}{N} \sum_{j,k} \Big(\langle \mathcal{T} [ \chi_{Rj}(u_2) &\chi_{Sj}(u_1) \chi_{Rk}(u_0) \chi_{Lk}(u_0) ]\rangle - \\
    &\langle \chi_{Rk}(u_0) \chi_{Lk}(u_0) \mathcal{T} [\chi_{Rj}(u_2) \chi_{Sj}(u_1)]\rangle\Big)
  \end{split} \\
  &\equiv - \mathcal{F}_{RS}(u_2, u_1, u_0)
\end{align}
where we have explicitly indicated which operators are contour-ordered by the symbol $\mathcal{T}$, and $S \in \{R, L\}$ indicates the side of the fermion insertion at $u_1$. By choosing the appropriate function $g_{R}$ or $g_{L}$, we can take the two operators $\chi_j$ on the same or opposite sides. Since the two-point function is determined by the complex reparametrization $\psi(u)$, the four-point function above can be obtained from perturbation theory to the dynamical system of $p,\phi$ discussed in Section~\ref{sec:response}.

Using the results of Section~\ref{sec:response}, we can give explicit expressions for the four-point function. Define
\begin{equation}
  A(u, u_0) = \frac{d \ln \psi'(u)}{dp(u_0)}, \; B(u, u_0) = \frac{d \psi(u)}{dp(u_0)},\label{eq:defAB app}
\end{equation}
then the two four-point functions are
\begin{align}
  \mathcal{F}_{RS}(u_2, u_1, u_0)
  &= (A(u_2, u_0) + A(u_1, u_0)^{*})
                          + \left( B(u_2, u_0) - B(u_1, u_0)^{*} \right) F_S(u_2,u_1) \\
  F_L(u_2, u_1)
  &= \tan \frac{\psi(u_2) - \psi(u_1)^{*}}{2} = \tan \left(  \frac{t_G(u_2) - t_G(u_1)}{2} - \frac{i}{2} \gamma(u_2, u_1) \right) \\
  F_R &= - 1 / F_L \\
  \label{eq:gamma-regulator-definition-app}
  \gamma(u_2, u_1) &= \tanh^{-1} \left( e^{- \phi(u_1)} \frac{1 + e^{- (\phi(u_2) - \phi(u_1))}}{1 + e^{- (\phi(u_2) + \phi(u_1))}} \right).
\end{align}

Explicit formulas for $A$ and $B$ in terms of $\tilde{\xi}(u)$ are
\begin{align}
  \label{eq:4pt-A-coeff-xsi}
  A(u, u_0)
  &= - \sqrt{\frac{1 - e^{- 2 \phi(u_0)}}{1 - e^{- 2 \phi(u)}}}
    \begin{pmatrix}
      1 \\ i
    \end{pmatrix}^{\dagger}
          \tilde{\xi}(u) \\
  \label{eq:4pt-B-coeff-xsi}
  B(u, u_0)
  &=
    - \frac{\sqrt{1 - e^{- 2 \phi(u_0)}}}{2}
    \left( \begin{pmatrix} 1 \\ i \end{pmatrix}^{\dagger}
                        \int_{u_0}^{u} e^{i p(u')} \csch \phi(u') \tilde{\xi}(u') du' \right).
\end{align}
We note that there is a relation between $A$ and $B$ from the fact that, with $u > u_0$, $\mathcal{F}_{RR}(u, u, u_0) \propto \sum_{jk} \langle [\chi_{Rj}(u) \chi_{Rj}(u), \chi_{Rk}(u_0) \chi_{Lk}(u_0)]\rangle = 0$,
\begin{equation}
  \label{eq:im-b-from-a}
  \Im B(u,u_0) = - e^{- \phi(u)} \Re A(u, u_0).
\end{equation}
From the right-left correlator, we have
\begin{equation}
  \label{eq:4pt-as-phi-response-probe}
  \frac{d \phi(u)}{dp(u_0)} = \{\phi(u_0), \phi(u)\} = - \frac{1}{2} \mathcal{F}_{RL} (u, u, u_0) = - (1 - e^{- 2 \phi(u)}) \Re A(u, u_0).
\end{equation}

Thus to find the four-point function, we need $G_1(u, u')$ to determine $A$ and the integral
\begin{equation}
  I_{B}(u, u', u_0) =
  \int_{u_0}^u \begin{pmatrix} \cos(p(v)) \\ \sin(p(v)) \end{pmatrix}^{\dagger} \csch \phi(v) G_1(v, u') dv
\end{equation}
for some value of $u'$ to determine $\Re B$. As described in Section~\ref{sec:response}, the point of defining the matrix $G_1$ is that we can ``patch together'' solutions in different time intervals by $G_1(u, u') = G_1(u, u'') G_1(u'', u')$. Likewise, the point of defining $I_B$ is that we can ``patch together'' solutions in a similar way: for example,
\begin{equation}
  \Re B(u, u') = - \frac{\sqrt{1 - e^{- 2 \phi(u')}}}{2}
  \left( I_B(u, u_a, u_c) \tilde{\xi}(u_a) + I_B(u_c, u_b, u') \tilde{\xi}(u_b) \right).
\end{equation}

As an example, for the fixed point solution we have
\begin{equation}
  I_{B}^{(G)}(u, u_0, u_0) =
  (1 + \coth \phi_G)
  \begin{pmatrix}
    \sqrt{\frac{2}{3 + \coth \phi_G}} \sin (\omega_G (u - u_0)) \\
    \frac{2}{\omega_G \sqrt{2 - e^{- 2 \phi_G}}} \sin^2 \frac{\omega_G(u - u_0)}{2}
  \end{pmatrix}.
\end{equation}
In the decoupled case,
\begin{equation}
  I_B^{(R)}(u, 0, u_0) = \left. 2 \Re \left[ \psi'(u) \begin{pmatrix} u \cos \epsilon - i \\ - \csc^2 \epsilon \end{pmatrix}^{T} \right] \right|_{u_0}^{u}.
\end{equation}
For later convenience, we also evaluate the Rindler contribution directly
\begin{multline}
  \Re B^{(R)}(u, u_0) = - \frac{1}{2}
  \Big[ \cot^2 \epsilon (\csch \phi(u_0) - \csch \phi(u) (1 + (\hat{u} - \hat{u}_0) \tanh \hat{u}_0)) \\ +  \tanh \hat{u}_0  (\csch \phi(u_0) \tanh \hat{u}_0 - \csch \phi(u) \tanh \hat{u}) \Big].
\end{multline}

\subsection{Four-point function in the rescued black hole geometry}\label{app:rescue 4pt}

For simplicity, we will consider the regime where the late rescue condition \eqref{eq:late rescue limit} holds and $\delta \phi > 0$. The only unknown ingredient is the matrix $G_1(u, u_{r-})$, defined in \eqref{eq:G1-matrix-def}, throughout the rescue region $u_{r-} < u \le u_{r+}$.

Recall that we are in the regime $e^{2 \phi_{r-}} \gg 1$ and $\delta \phi > 0$, where we can do perturbation theory in $e^{- 2 \phi}$. Define
\begin{equation}
  \label{eq:T_r-definition}
  T_{r}(u) = \int_{u_{r-}}^{u} \frac{\mu \phi - \lgqeham}{1 - e^{- 2 \phi}}, \; \lambda_T = T_{r}(u_+) / (u_{r+} - u_{r-})
\end{equation}
then to lowest order in $e^{- 2 \phi}$ we have
\begin{equation}
  G^{(r)}_1(u, u_{r-}) \approx e^{T_{r}(u) B_+} = 1 + T_{r}(u) B_+.
\end{equation}
To get an order of magnitude estimation, it is useful to keep in mind that $\lambda_T \sim O(1)$, and when $|p_{r-}| \ll 1$ we have $\lambda_T \approx 1$. For convenience, we will just assume $|p_{r-}| \ll 1$ and estimate
\begin{equation}
  I_B^{(r)}(u_{r+}, u_{r-}, u_{r-}) \sim
  \begin{pmatrix}
    \cos p_{r-} \\ \sin p_{r-}
  \end{pmatrix}^{\dagger}
  \csch \phi_{r-} (u_{r+} - u_{r-}).
\end{equation}

Now we can find the response $\tilde{\xi}(u)$ by propagating by $G_1$ on the appropriate interval. For example, if $u_0 < u_{r-} < u_{r+} < u$, we have
\begin{equation}
  \tilde{\xi}(u) = G_1^{(G)}(u, u_{r+}) G^{(r)}_1(u_{r+}, u_{r-}) G^{(R)}_1(u_{r-}, u_0)\xi(u_0).
\end{equation}
We find $A$, and hence $\Im B$, from \eqref{eq:4pt-A-coeff-xsi}. We use the integrals $I_B$ (and the prefactor from \eqref{eq:4pt-B-coeff-xsi}) to find $\Re B$. For example, with the same times $u_0$ and $u$
\begin{multline}
  \label{eq:re-b-from-IB-nontrivial-example}
  \Re B(u, u_0) = \\ - \frac{\sqrt{1 - e^{- 2 \phi(u_0)}}}{2} \left( I^{(G)}_B(u, u_{r+}, u_{r+}) \tilde{\xi}(u_{r+}) + I^{(r)}_B(u_{r+}, u_{r-}, u_{r-}) \tilde{\xi}(u_{r-})
  \right) \\ + \Re B^{(R)}(u_{r-}, u_0).
\end{multline}

As an example, we find the full four-point function $\mathcal{F}_{RL}(u, u, u_0)$ in the rescued geometry when $u_0 < u_{r-} < u_{r+} < u$, to the leading order in low initial temperature $\epsilon \ll 1$ and late enough rescue (such that \eqref{eq:late rescue limit} holds)
\begin{multline}
  \label{eq:approximate-sensitivity-4pt}
  \mathcal{F}_{RL}(u, u, u_0)
  \approx 2 \cot \epsilon \\
  \Bigg\{
  \cos(\eta(\mu_2) \sqrt{2} (t_G(u) - t_G(u_{r+})))
  \Big[\tanh(\hat{u}_{r-}) - \tanh(\hat{u}_0) + (\hat{u}_{r-} - \hat{u}_0) \tanh(\hat{u}_{r-}) \tanh(\hat{u}_0)\Big] \\
  + \left(T_r(u_{r+}) \cos(\eta(\mu_2) \sqrt{2} (t_G(u) - t_G(u_{r+}))) + e^{\phi(u_{r+})} \sin(\eta(\mu_2) \sqrt{2} (t_G(u) - t_G(u_{r+})))\right) \\ \Big[\tan \epsilon \tanh(\hat{u}_{r-}) \tanh(\hat{u}_0) + \cot (\epsilon) e^{- 2 \phi(u_{r-})} (1 + (\hat{u}_{r-} - \hat{u}_0) \tanh (\hat{u}_0))\Big]
  \Bigg\}.
\end{multline}
where we have defined the factor
\begin{equation}
  \label{eq:eta-trig-arg-factor}
  \eta(\mu) = \frac{\omega_G(\mu)}{\sqrt{2} V_G(\mu)}
  = \sqrt{1 - e^{- 2 \phi_G(\mu)}} \sqrt{\frac{3 + \coth \phi_G(\mu)}{4}},
\end{equation}
$T_r(u_{r+})$ is defined and discussed near \eqref{eq:T_r-definition}, and kept the notation $\hat{u} \equiv \sin \epsilon u$ from Section~\ref{sec:response}. As mentioned above, $T_r(u_{r+}) / (u_{r+} - u_{r-}) \sim O(1)$ and depends on the details of the rescue. By the relation \eqref{eq:4pt-as-phi-response-probe} this also gives an expression for the $A(u, u_0)$ coefficient in this setup, as well as the imaginary part of $B(u, u_0)$ by \eqref{eq:im-b-from-a}. At late rescue time, $\phi(u_{r+})$ grows linearly with $u_{r+}$, and the term with $e^{\phi(u_{r+})}$ in the second line of Eq. \eqref{eq:approximate-sensitivity-4pt} becomes the dominant term that grows exponentially with time (except when the oscillating $\sin$ function is close to zero). Keeping this term leads to Eq. (\ref{eq:FRLuuu0 rescue}).

\end{document}
